\documentclass[a4paper,10pt]{article}
\usepackage{amsmath,amssymb,leftidx,cite,epsfig}
\usepackage[utf8x]{inputenc}

\title{ Poincaré Invariant Quantum Field Theories With Twisted Internal Symmetries}
\author{Rahul Srivastava\footnote {rahul@cts.iisc.ernet.in} \, and Sachindeo Vaidya\footnote {vaidya@cts.iisc.ernet.in}\\
\\
\begin{small}
 Centre for High Energy Physics, Indian Institute of Science, Bangalore, India
\end{small}}

\begin{document}

\date{\empty}
\maketitle

\begin{abstract}
Following up the work of \cite{bal-amilcar} on deformed algebras, we present a class of Poincaré invariant quantum field theories with particles having deformed internal symmetries. 
The twisted quantum fields discussed in this work satisfy commutation relations different from the usual bosonic/fermionic 
commutation relations. Such twisted fields by construction are nonlocal in nature. Despite this nonlocality we show that   
it is possible to construct interaction Hamiltonians which satisfy cluster decomposition principle and are Lorentz invariant. We further illustrate these ideas by considering
global $SU(N)$ symmetries. Specifically we show that twisted internal symmetries can provide a natural framework for the discussion of the marginal deformations ($\beta$-deformations)
of the $\mathcal{N}=4$ SUSY theories.  
\end{abstract}

\vspace{2mm}


The deformation of spacetime symmetries leading to twisted statistics for quantum fields is not a special feature of only spacetime symmetries \cite{Mack:1991tg,mack3}. 
In general there exists a way of deforming the algebra of functions on a manifold $M$ which can lead to deformed statistics in a quantum theory \cite{bal-amilcar}. Motivated by this, 
in this paper we discuss the possibility of having twisted statistics by deforming global internal symmetries. Following up the work of \cite{bal-amilcar} on deformed algebras, 
we present a class of Poincaré invariant quantum field theories with twisted bosonic/fermionic particles having
deformed internal symmetries. In other words, we construct field theories where 
fields transform in standard way under Poincar\'e transformation (as opposed to the twisted transformation on noncommutative spacetimes e.g. Groenewold-Moyal (GM) plane)
 but nonetheless have twisted statistics. 
The twisted quantum fields discussed in this work, satisfy commutation relations different from the usual bosonic/fermionic 
commutation relations. Such twisted fields by construction (and in view of CPT theorem) are nonlocal in nature. We show that inspite of the basic ingredient fields being nonlocal,  
it is possible to construct interaction Hamiltonians which satisfy cluster decomposition principle and are Poincaré invariant. Although the formalism developed here can be
adapted to the discussion of a generic global internal symmetry group but for sake of concreteness we restrict ourself only to the discussion of global $SU(N)$ symmetries. 
As a specific example of interesting application of these ideas we show that twisted internal symmetries provide a natural framework for the discussion of the marginal deformations
(also known as $\beta$-deformations) of the scalar sector of $\mathcal{N}=4$ supersymmetric (SUSY) theories.

As we will elaborate in sections [2] and [3], the twisted field theories discussed here are characterized by the 
``twist element'' $e^{\frac{i}{2}Q_l \theta_{lm} Q_m}$, $Q_l$ being the Cartan charge of the $SU(N)$ group and the $\theta$ matrix consisting of arbitrary dimensionless parameters. 
The twist element determines the action of $SU(N)$ on multi-particle states \cite{Mack:1991tg,mack3} 
and as we will show, the action of the $SU(N)$ group elements on multi-particle states gets changed.
Hence, all the new effects discussed in this work arise due to the twist element. In the limit of $\theta_{lm} \rightarrow 0$, one recovers back the usual group theoretic results.  

In which regime of investigation would the effects of this twist be observable? In the case of GM plane, 
the answer is intuitively straightforward: since the twist is of the form $e^{\frac{i}{2} \theta_{\mu \nu} p^\mu 
p^\nu}$, the $\theta$ matrix has dimensions of $(\rm{length})^2$. The twist element differs appreciably from 1 only
when the momenta for the process(es) at hand are comparable to the inverse length scale inherent in 
$\theta$. For instance in situations involving (quantum) gravity, this scale is taken to be the Planck length $l_P$, i.e. 
$\theta_{\mu \nu} = \theta^{(0)}_{\mu \nu} l_P^2$, where $\theta^0_{\mu\nu}$ are dimensionless numbers of 
order 1. In physics of the quantum Hall effect, this length scale is the magnetic length $l_B = 1/\sqrt{q B}$.

When could the twist $e^{\frac{i}{2}Q_l \theta_{lm} Q_m}$ be appreciably different from 1? Since 
the internal charges are typically dimensionless, the argument is a little more subtle here. To 
answer this question, let us re-write the exponent $\theta_{lm} Q_l Q_m$ as $\theta^{(0)}_{lm} \alpha \lambda_l 
\lambda_m$, where the $\lambda$'s are the eigenvalues of the charge operator, $\theta^{(0)}_{lm}$ are dimensionless numbers of order 1, and $\alpha$ is the 
coupling constant of the theory. For instance $\alpha = \alpha_e$, the fine structure constant in case of 
QED, and $\alpha = \alpha_s$ for QCD. Due to renormalization effects (assuming that the symmetry remains unbroken both perturbatively as well as non-perturbatively), 
the $\alpha$ generally runs as a function of the centre-of-mass energy $\sqrt{s}$. 

In the regime where perturbation theory is valid, i.e. $\alpha$ is small, the twist factor is thus small 
and produces little or no observable effects. The situation, however can change substantially in regimes 
where $\alpha \simeq 1$. The twist can now differ appreciably from 1. For the case of QED, this regime is 
in the disappointingly distant ultraviolet where $\alpha_e \simeq 1$. For QCD, the situation is much better: it is 
at low energies that $\alpha_s$ approaches 1. One may expect the effect of the twist to make itself felt 
as one approaches $\Lambda_{QCD}$ from the ultraviolet.

The above argument, though heuristic, is quite suggestive of the regime where the effects of the twist may be 
manifested. It is of course possible that the $\theta^{(0)}_{lm}$ are (very close to) zero rather than 1, but 
this can only be determined by experiments.

The plan of the paper is as follows.  We begin with briefly reviewing the treatment of global symmetries, in particular $SU(N)$ group, in the usual untwisted case. 
We then discuss a specific type of twist called ``antisymmetric twist''. 
 This kind of twist is quite similar in spirit to the twisted noncommutative field theories  and is parameterized by an antisymmetric matrix $\theta$. 
The formalism developed here will closely resemble (with generalizations and modifications which we will elaborate on) the formalism of twisted noncommutative theories 
\cite{bal-pinzul}. We then construct interaction terms using these twisted fields and discuss the scattering problem for such theories. 
We show that the twisted $SU(N)$ invariant interaction Hamiltonian as well as the $S$-matrix elements are identical to their untwisted analogues and hence by doing a 
scattering experiment it is difficult to distinguish between a twisted and untwisted theory. We further show that relaxing the demand of $SU(N)$ invariance leads to 
differences between the two theories and for certain interaction terms the twisted theory is nonlocal although its analogous untwisted theory is still local. 
As an interesting example of twisted field theories we also show that the marginal deformations of the scalar sector (without gauge fields) of $\mathcal{N}=4$ SUSY interaction Hamiltonian 
density can be mapped to a twisted interaction Hamiltonian density. 
 
We then construct more general twisted statistics which can be viewed as internal symmetry analogue of dipole theories which arise in the low energy limit of 
certain string configurations \cite{dipole-keshav}. We also discuss the construction of interaction 
terms and scattering formalism for it. The main results for general twists are same as those for the antisymmetric twist. 

We end the work with discussion of causality of the twisted field theories. We show that the twisted fields are noncausal and hence a generic observable constructed out 
of them is also noncausal. Inspite of this it is possible to construct certain interaction Hamiltonians, e.g. the $SU(N)$ invariant interaction Hamiltonian, which are causal and 
satisfy cluster decomposition principle.


 \section{Brief Review of Global Symmetries In Untwisted Case}


In this section we briefly review the standard treatment of global symmetries in the usual untwisted case \cite{weinberg, greiner, greiner-sym, ait}. In view of the computations
 and generalizations done in later parts, 
we take the route of Hamiltonian formalism, instead of the more convenient Lagrangian formalism, to study the global symmetries. For sake of simplicity, we mostly 
restrict our discussion to the case of matter fields which transform as scalars under Poincaré group and transform as a fundamental representation of a given 
global $SU(N)$ symmetry group. The treatment here can be easily generalized to spinor fields, as well as, higher representations of $SU(N)$ group. 

Let $\phi_r (x)$, $r = 1, 2, \dots N $ be a set of complex scalar (under Lorentz transformation) quantum fields having mode expansion given by\footnote{ Throughout this work we 
will denote the usual bosonic/fermionic annihilation operators by the labels $c_r$ and $d_r$ whereas the twisted operators will be denoted by $a_r$ and $b_r$. The same notation will 
be followed for usual and twisted creation operators.}
\begin{eqnarray}
\phi_r (x) & = & \int \frac{d^3 p}{(2\pi)^3} \frac{1}{2E_{p}} \left[ c_r(p) e^{-i px} \, + \, d^{\dagger}_r(p) e^{i px} \right ], \nonumber \\
\phi^\dagger_r (x) & = & \int \frac{d^3 p}{(2\pi)^3} \frac{1}{2E_{p}} \left[ d_r(p) e^{-i px} \, + \, c^{\dagger}_r(p) e^{i px} \right ].
\label{modeexpansioncom}
\end{eqnarray}
The creation/annihilation operators satisfy the commutation relations 
\begin{eqnarray}
  c^\#_r(p_{1}) c^\#_s(p_{2}) & = & \eta \, c^\#_s(p_{2}) c^\#_r(p_{1}), \nonumber \\
(c^{\dagger}_r)^\# (p_{1}) (c^{\dagger}_s)^\# (p_{2}) & = & \eta \,(c^{\dagger}_s)^\# (p_{2})(c^{\dagger}_r)^\# (p_{1}), \nonumber \\
c_r (p_{1})  c^{\dagger}_s(p_{2}) & = & \eta \,  c^{\dagger}_s (p_{2}) c_r (p_{1}) + (2\pi)^3 \, 2 E_p \, \delta_{rs} \, \delta^{3}(p_{1} \, - \, p_{2}), \nonumber \\
d_r (p_{1}) d^{\dagger}_s (p_{2}) & = & \eta \,  d^{\dagger}_s (p_{2}) d_r (p_{1}) + (2\pi)^3 \, 2 E_p \, \delta_{rs} \, \delta^{3}(p_{1} \, - \, p_{2}), \nonumber \\
c_r (p_{1}) d^{\dagger}_s (p_{2}) & = & \eta \, d^{\dagger}_s (p_{2}) c_r (p_{1}), 
\label{cstcom}
\end{eqnarray}
where $c^\#_r ({p})$ stands for either of the operators $c_r (p), d_r (p)$ and $(c^{\dagger}_r)^\# (p) $ stands for either of the operator $c^\dagger_r (p), d^\dagger_r (p)$. 
Also, as these 
are bosonic operators, so $\eta = 1$ should be taken in (\ref{cstcom}). For the case of fermionic operators $\eta = -1$ should be taken. Since we are mostly concerned with 
internal symmetries, so we will usually suppress the momentum dependence of the operators and will not write them explicitly unless we need them.  

Let $U(\sigma) = \exp(i \sigma_a \Lambda_a)$, $a = 1, 2, \dots (N^2 - 1) $ and $\sigma$ being $(N^2 - 1) $ arbitrary parameters (independent of spacetime coordinates), be the unitary 
representation of the group element ``$\sigma$'' of the $SU(N)$ group on Fock space. Then, if the fields $\phi_r (x)$ transform as fundamental representation of the $SU(N)$, we have
\begin{eqnarray}
 U(\sigma) \phi_r (x) U^\dagger(\sigma) & = &  \phi'_r (x) \, = \, \left( e^{-i \sigma_a T_a} \right)_{rs} \,\phi_s (x), \nonumber \\
  U(\sigma) \phi^\dagger_r (x) U^\dagger(\sigma) & = &  \phi'^\dagger_r (x) \, = \, \left( e^{i \sigma_a T_a} \right)_{sr} \,\phi^\dagger_s (x), 
\label{suntranformcom}
\end{eqnarray}
where $r,s = 1, \dots N$ and $ a = 1, \dots (N^2 - 1)$. Also, $T_a$ are $N \times N $  hermitian matrices and furnish the fundamental representation of the generators of the 
group satisfying the Lie algebra
\begin{eqnarray}
 [T_a, T_b] & = & i f_{abc}\, T_c
\label{sunlie}
\end{eqnarray}
where $f_{abc}$ are the structure constants. 

The infinitesimal version of (\ref{suntranformcom}) can be written as 
\begin{eqnarray}
 U(\epsilon) \phi_r (x) U^\dagger(\epsilon) & = &  \phi'_r (x) \, = \, \phi_r(x) - i \epsilon_{a} (T_{a})_{rs} \phi_s(x), \nonumber \\
U(\epsilon) \phi^\dagger_r (x) U^\dagger(\epsilon) & = &  \phi'^\dagger_r (x) \, = \, \phi^\dagger_{r}(x) + i \epsilon_{a} (T_{a})_{sr} \phi^\dagger_s(x).
\label{infitsuntrans}
\end{eqnarray}
The $\Lambda_a$ furnish a Fock space representation of the generators of the group, having the form
\begin{eqnarray}
 \Lambda_a & = &  \int \frac{d^3 p}{(2\pi)^3} \frac{1}{2E_{p}} \left[ (T_a)_{rs} c^\dagger_r c_s \, - \, (T_a)^\ast_{rs}  d^\dagger_r d_s \right],
\label{fockgenratorssun}
\end{eqnarray}
where $(-T^\ast_a) = (-T^T_a)$ are $N \times N $ matrices  \footnote{$T^T_a$ stands for transpose of $T_a$. Also this relation holds because of hermiticity of the generators.}
and furnish the anti-fundamental representation of the generators of the group satisfying the Lie algebra
\begin{eqnarray}
 [(-T^\ast_a), (-T^\ast_b)] & = & i f_{abc}\, (-T^\ast_c).
\label{antisunlie}
\end{eqnarray}
The operators $\Lambda_a$ also satisfy the same Lie algebra
\begin{eqnarray}
 [\Lambda_a, \Lambda_b] & = & i f_{abc}\, \Lambda_c.
\label{focksunlie}
\end{eqnarray}  
Using (\ref{fockgenratorssun}) one can immediately check the correctness of (\ref{suntranformcom}) and can deduce the transformation properties of the 
creation/annihilation operators which are given by
\begin{eqnarray}
 U(\sigma) c_r U^\dagger(\sigma) & = &  c'_r \, = \, \left( e^{-i \sigma_a T_a} \right)_{rs} \,c_s, \nonumber \\
 U(\sigma) c^\dagger_r U^\dagger(\sigma) & = &  c'^\dagger_r \, = \, \left( e^{i \sigma_a T_a} \right)_{sr} \,c^\dagger_s, \nonumber \\
U(\sigma) d_r U^\dagger(\sigma) & = &  d'_r \, = \, \left( e^{i \sigma_a T^\ast_a} \right)_{rs} \,d_s, \nonumber \\
 U(\sigma) d^\dagger_r U^\dagger(\sigma) & = &  d'^\dagger_r \, = \, \left( e^{-i \sigma_a T^\ast_a} \right)_{sr} \,d^\dagger_s.
\label{creanisuntranformcom}
\end{eqnarray}
Using (\ref{creanisuntranformcom}) and assuming that vacuum remains invariant under the transformations i.e. $ U(\sigma) | 0 \rangle = | 0 \rangle$, we can  
deduce the transformation property of state vectors, which for single-particle states, is given by
\begin{eqnarray}
 U(\sigma) | r \rangle & = &  U(\sigma) c^\dagger_r | 0 \rangle \, = \, U(\sigma) c^\dagger_r U^\dagger(\sigma) U(\sigma) | 0 \rangle 
\, = \, \left( e^{i \sigma_a T_a} \right)_{sr} \,c^\dagger_s | 0 \rangle \, = \, \left( e^{i \sigma_a T_a} \right)_{sr}  | s \rangle, \nonumber \\
 U(\sigma) \overline{| r \rangle} & = &  U(\sigma) d^\dagger_r | 0 \rangle \, = \, U(\sigma) d^\dagger_r U^\dagger(\sigma) U(\sigma) | 0 \rangle 
\, = \, \left( e^{-i \sigma_a T^\ast_a} \right)_{sr} \,d^\dagger_s | 0 \rangle \, = \, \left( e^{-i \sigma_a T^\ast_a} \right)_{sr}  \overline{| s \rangle}.
\label{statevecsuntransform}
\end{eqnarray}
Similar transformation properties hold for multi-particle states. 

Having discussed the transformation properties of quantum fields and state vectors under the $SU(N)$ group, let us now consider a Hamiltonian density $\mathcal{H} (x)$ constructed 
out of the fields $\phi_r (x)$ and their canonical conjugates $\Pi_r (x)$. The operator $U(\sigma) = \exp(i \sigma_a \Lambda_a)$ transforms it as
\begin{eqnarray}
 U(\sigma) \mathcal{H} (x) U^\dagger(\sigma) & = & \mathcal{H}' (x).
\label{Hamiltoniansun}
\end{eqnarray}
The transformation of (\ref{Hamiltoniansun}) is said to be a symmetry transformation and the system is said to be having a $SU(N)$ global symmetry if the Hamiltonian density remains 
invariant under such a transformation\footnote{Since these are global transformations so it is enough for $\mathcal{H} (x)$ to be invariant, which will automatically imply 
that the Hamiltonian $H$ itself remains invariant.}. Therefore we have 
\begin{eqnarray}
 U(\sigma) \mathcal{H} (x) U^\dagger(\sigma) & = & \mathcal{H}' (x) \, = \, \mathcal{H} (x). 
\label{hamsun}
\end{eqnarray}
The above condition in turn implies that 
\begin{eqnarray}
 [H, \Lambda_a] & = & 0.
\label{symmofh}
\end{eqnarray}
From (\ref{symmofh}) we can infer that all $\Lambda_a$ are constants of motion and hence conserved quantities, called `` charge operators'' and their eigenvalues are termed as
 ``internal charges'' of the given eigenstate. Since $SU(N)$ is a nonabelian group satisfying the Lie algebra (\ref{focksunlie}), a state cannot simultaneously be 
an eigenstate of all the charge operators and hence only a subset of the charges can be simultaneously measured.  The maximal commuting subset of the charge operators is called 
 ``Cartans'' of the group and usually the eigenstates of the Cartans are 
taken as the basis states. For a $SU(N)$ group there are $N-1$ Cartans and we will denote them by $Q_m$; $1 \leq m \leq N-1 $. 

The condition (\ref{hamsun}) puts stringent constraints on the type of Hamiltonian densities allowed by the symmetry. For example, a free theory Hamiltonian density 
will satisfy (\ref{hamsun}) if and only if the masses $m_r$ of all particles are same i.e. $m_r = m_s = \cdots  = m$ and is given by
 \begin{eqnarray}
  \mathcal{H}_0 & = &  \Pi^{\dagger}_r \Pi_r  \, + \, (\partial_i \phi^{\dagger}_r) (\partial^i \phi_r) \, + \, m^2  \, \phi^{\dagger}_r \phi_r.
\label{freeham}
 \end{eqnarray}
where $\Pi_{r}$ is the canonical conjugate momentum.
The only renormalizable interaction Hamiltonian density compatible with (\ref{hamsun}) is given by
\begin{eqnarray}
 \mathcal{H}_{\rm{int}} & = &  \frac{\gamma}{4} \, \phi^{\dagger}_r \phi^{\dagger}_s \phi_r \phi_s,
\label{gaugen4susy}
\end{eqnarray}
where $r,s = 1, 2, \cdots, N$.


\subsection*{Weight Basis}


The discussion till now holds for generators written in any basis. For purpose of our work, it is convenient to write them in ``weight basis''. From now onwards we will write the 
generators in weight basis only. The obvious advantage of working in weight basis being that the Cartans are all diagonal matrices and easy to deal with.  
In this basis, we denote Cartans by $Q_m$; $1 \leq m \leq N-1 $. The other generators, denoted by $E_n$; $1 \leq n \leq N(N-1) $, are the so called
``raising/lowering'' generators. Their Fock space representation is given by  
\begin{eqnarray}
 Q_m & = &  \int \frac{d^3 p}{(2\pi)^3} \frac{1}{2E_{p}} \left[ (q_m)_{rs} c^\dagger_r c_s \, - \, (q_m)^\ast_{rs}  d^\dagger_r d_s \right], \nonumber \\
 E_n & = &  \int \frac{d^3 p}{(2\pi)^3} \frac{1}{2E_{p}} \left[ (e_n)_{rs} c^\dagger_r c_s \, - \, (e_n)^\ast_{rs}  d^\dagger_r d_s \right],
\label{fockcargensun}
\end{eqnarray}
where  creation/annihilation operators in (\ref{fockcargensun}) are labeled using weights. Also, $q_m$ and $e_n$ are $N \times N $ matrices and they together satisfy the lie 
algebra (\ref{sunlie}) of the group, furnishing the fundamental representation of the generators. 
The $N \times N $ matrices $(-q^\ast_m)$ and $(-e^\ast_n)$ furnish the anti-fundamental representation of the generators. Moreover since $q_m$ and $q^\ast_m$ are Cartans of the group 
and are written in the weight basis so
\begin{eqnarray}
[q_m , q_{m'}] & = & 0,
\label{comCartan}
\end{eqnarray}
and
\begin{eqnarray}
q_m  \, = \,q^\ast_m & = &  
  \begin{pmatrix}
  \lambda_m^1 & 0 & \cdots & 0 \\
  0 & \lambda_m^2 & \cdots & 0 \\
  \vdots & \vdots   & \ddots & \vdots \\
  0 & 0 & \cdots & \lambda_m^{N}
 \end{pmatrix}.
\label{digCartan}
\end{eqnarray}
In the weight basis we have 
\begin{eqnarray}
 \left[Q_m , c^{\dagger}_u \right] & = &  \lambda_{m}^{(u)} c^{\dagger}_u,
\label{weightcre}
\end{eqnarray}
where $\lambda_{m}^{(u)}$ is the mth component of the weight vector corresponding to the Cartan $Q_m$. 

Similarly we have
\begin{eqnarray}
 \left[Q_m , c_u \right] & = & - \lambda_{m}^{(u)} c_u, \nonumber \\
\left[Q_m , d_u \right] & = &   \lambda_{m}^{(u)} d_u, \nonumber \\
\left[Q_m , d^{\dagger}_u \right] & = &  - \lambda_{m}^{(u)} d^\dagger_u.
\label{weightani}
\end{eqnarray}
Using (\ref{weightani}) and (\ref{weightcre}) we have 
\begin{eqnarray}
 \left[Q_m , \phi_u \right] & = & - \lambda_{m}^{(u)} \phi_u, \nonumber \\
\left[Q_m , \phi^{\dagger}_u \right] & = &   \lambda_{m}^{(u)} \phi^\dagger_u.
\label{chargefields}
\end{eqnarray}
The above was a very brief review of the standard discussion of global symmetries in quantum field theories. Apart from completeness of the work, the main purpose of this section 
was to setup the notations and conventions that we use in the rest of the paper. Keeping that in mind we restricted ourself to the discussion of only $SU(N)$ group symmetries
and to only scalar fields transforming as a fundamental representation. Other type of groups like $SO(N)$ can be discussed in a similar way. Also, within $SU(N)$
group symmetries, generalization to higher representations and discussion of transformation properties of spinor (under Lorentz transformation) fields 
can be done in a similar and straightforward way. 


\section{Antisymmetric Twists}


Our main interest is to write field theories where the particles satisfy twisted commutation relations, which in general can be 
\begin{eqnarray}
a_r ({p_{1}}) a_s ({p_{2}}) & = & \zeta_1  \,a_s ({p_{2}}) a_r ({p_{1}}), \nonumber \\
a^{\dagger}_r ({p_{1}}) a^{\dagger}_s ({p_{2}}) & = & \zeta_2 \, a^{\dagger}_s ({p_{2}}) a^{\dagger}_r ({p_{1}}), \nonumber \\
\vdots \nonumber \\
a_r ({p_{1}})  a^{\dagger}_s ({p_{2}})  & = & \zeta_3 \,  a^{\dagger}_s ({p_{2}}) a_r ({p_{1}}) + (2\pi)^3 \, 2 E_p \, \delta_{rs} \, \delta^{3}(p_{1} \, - \, p_{2}), \nonumber \\
b_r ({p_{1}}) b^{\dagger}_s ({p_{2}}) & = & \zeta_4 \, b^{\dagger}_s ({p_{2}}) b_r ({p_{1}}) + (2\pi)^3 \, 2 E_p \, \delta_{rs} \, \delta^{3}(p_{1} \, - \, p_{2}), \nonumber \\
a_r ({p_{1}}) b^{\dagger}_s ({p_{2}}) & = & \zeta_5 \, b^{\dagger}_s ({p_{2}}) a_r ({p_{1}}),  
\label{gtwistedcomrel}
\end{eqnarray}
where we are denoting the twisted creation and annihilation operators for particles and anti-particles by $a_r, b_r$ and $a^\dagger_r, b^\dagger_r$ respectively. 

In this section, we restrict ourself only to the discussion of  a specific type of twist which we call ``antisymmetric twist''. This kind of twist is quite similar in spirit to the 
twisted statistics of noncommutative field theories e.g Groenewold-Moyal (GM) plane. The formalism developed here will closely resemble (with appropriate generalizations 
and modifications) the formalism of twisted noncommutative theories \cite{bal-pinzul}. Also, the formalism developed here is true for any $SU(N)$ group with N $\geq$ 3. 
The antisymmetric twisted commutation relations for $a_r, b_r$ and $a^\dagger_r, b^\dagger_r$ are
\begin{eqnarray}
 a^\# _r ({p_{1}}) a^\#_s ({p_{2}}) & = & \eta \, e^{ i\tilde{\lambda}^{(r)} \wedge \tilde{\lambda}^{(s)}}  \, a^\#_s ({p_{2}}) a^\#_r ({p_{1}}), \nonumber \\
(a^{\dagger}_r)^\# ({p_{1}}) (a^{\dagger}_s)^\# ({p_{2}}) & = & \eta \,e^{ i\tilde{\lambda}^{(r)} \wedge \tilde{\lambda}^{(s)}} \, (a^{\dagger}_s)^\# ({p_{2}}) 
(a^{\dagger}_r)^\# ({p_{1}}), \nonumber \\
a_r ({p_{1}})  a^{\dagger}_s ({p_{2}})  & = & \eta \,e^{ - i\tilde{\lambda}^{(r)} \wedge \tilde{\lambda}^{(s)}}  \, a^{\dagger}_s ({p_{2}}) a_r ({p_{1}})
 + (2\pi)^3 \, 2 E_p \, \delta_{rs} \, \delta^{3}(p_{1} \, - \, p_{2}), \nonumber \\
b_r ({p_{1}}) b^{\dagger}_s ({p_{2}}) & = & \eta \,e^{ -i\tilde{\lambda}^{(r)} \wedge \tilde{\lambda}^{(s)}} \, b^{\dagger}_s ({p_{2}}) b_r ({p_{1}}) 
+ (2\pi)^3 \, 2 E_p \, \delta_{rs} \, \delta^{3}(p_{1} \, - \, p_{2}), \nonumber \\
a_r ({p_{1}}) b^{\dagger}_s ({p_{2}}) & = & \eta \, e^{ i\tilde{\lambda}^{(r)} \wedge \tilde{\lambda}^{(s)}} \, b^{\dagger}_s ({p_{2}}) a_r ({p_{1}}),  
\label{atwistedcomrel}
\end{eqnarray}
where  $\tilde{\lambda}^{(r)} \wedge \tilde{\lambda}^{(s)} = \tilde{\lambda}^{(r)}_{l} \tilde{\theta}_{lm} \tilde{\lambda}^{(s)}_{m} $; $l,m = 1, 2, \cdots, (N-1)$.
Right now $\tilde{\lambda}^{(r)}_{l}, \tilde{\lambda}^{(s)}_{m} $ are some arbitrary parameters whose meaning will be clarified soon. 
Also, $\tilde{\theta}_{lm} = - \tilde{\theta}_{ml}$  is an arbitrary 
real antisymmetric matrix. Moreover, $a^\#_r $ and $(a^{\dagger}_r)^\# $ stands for either of the operators $a_r, b_r$ and $a^\dagger_r, b^\dagger_r$ respectively. 
Again, for ``twisted bosons'' $\eta = 1$ and for ``twisted fermions'' $\eta = -1$ should be taken in (\ref{atwistedcomrel}).  

Using the above creation/annihilation operators we can write down complex scalar (under Lorentz transformations) quantum fields. Let $\phi_{\tilde{\theta}, r}(x)$, 
$r = 1, 2, \dots N $ be such a set of complex scalar quantum fields having mode expansion given by
\begin{eqnarray}
  \phi_{\tilde{\theta}, r} (x) & = & \int \frac{d^3 p}{(2\pi)^3} \frac{1}{2E_{p}} \left[ a_r(p) e^{-i px} \, + \, b^{\dagger}_r(p) e^{i px} \right ], \nonumber \\
\phi^\dagger_{\tilde{\theta}, r} (x) & = & \int \frac{d^3 p}{(2\pi)^3} \frac{1}{2E_{p}} \left[ b_r(p) e^{-i px} \, + \, a^{\dagger}_r(p) e^{i px} \right ].
\label{modeexpansionatwist}
\end{eqnarray}
The Fock space states can be similarly constructed using these twisted operators. To start with we assume that the vacuum of the twisted theory is same as that of the untwisted theory. 
The reason for the above assumption will be clarified soon. The multi-particle states can be obtained by acting the twisted creation operators on the vacuum state. 
Because of the twisted statistics (\ref{atwistedcomrel}), there is an ambiguity in defining the action of the twisted creation and annihilation operators on Fock space states.
We choose to define $a^\dagger_r (p)$, $p$ being the momentum label, to be an operator which adds a particle to the right of the particle list i.e.
\begin{eqnarray}
a^\dagger_r (p) | p_1,r_1; \, p_2,r_2;\, \dots \, p_n,r_n \rangle_{\tilde{\theta}} & = & | p_1,r_1;\, p_2,r_2;\, \dots \, p_n,r_n;\, p,r \rangle_{\tilde{\theta}}.
\label{actiontanticre}
\end{eqnarray}
It should be remarked that the particular choice (\ref{actiontanticre}) is purely conventional and we could have chosen the other convention where $a^\dagger_r (p)$
adds a particle to the left of the particle list. The two choices are not independent but are related to each other by a phase. Furthermore, the other choice can only result in 
an overall phase in the $S$-matrix elements.
With this convention, the single-particle Fock space states for this twisted theory are given by 
\begin{eqnarray}
\overline{|p, r \rangle}_{\tilde{\theta}} & = &   b^\dagger_r (p) | 0 \rangle,  \nonumber \\
|p, r \rangle_{\tilde{\theta}} & = &  a^\dagger_r (p) | 0 \rangle.
\label{sat}
\end{eqnarray}
The multi-particle states are given by
\begin{eqnarray}
\overline{| p_1, r_1; \, p_2,r_2; \, \dots \, p_n,r_n \rangle}_{\tilde{\theta}} & = &  b^\dagger_{r_n} (p_n) \dots b^\dagger_{r_2} (p_2) b^\dagger_{r_1} (p_1) | 0 \rangle, \nonumber \\
|p_1, r_1; \, p_2,r_2; \, \dots \, p_n,r_n \rangle_{\tilde{\theta}} & = &  a^\dagger_{r_n} (p_n) \dots a^\dagger_{r_2} (p_2) a^\dagger_{r_1} (p_1) | 0 \rangle .
\label{mat}
\end{eqnarray}
Owing to the twisted commutation relations of (\ref{atwistedcomrel}), the state vectors also satisfy a similar twisted relation e.g. for two-particle states we have 
\begin{eqnarray}
 \overline{| p_2, r_2; \, p_1,r_1 \rangle}_{\tilde{\theta}} & = & e^{ i\tilde{\lambda}^{(r_1)} \wedge \tilde{\lambda}^{(r_2)}} 
\, \overline{| p_1, r_1; \, p_2,r_2 \rangle}_{\tilde{\theta}}, \nonumber \\
 |p_2, r_2; \, p_1,r_1 \rangle_{\tilde{\theta}} & = &  e^{ i\tilde{\lambda}^{(r_1)} \wedge \tilde{\lambda}^{(r_2)}} \, |p_1, r_1; \, p_2,r_2 \rangle_{\tilde{\theta}}.  
\label{2statecom}
\end{eqnarray}


\subsection*{Dressing Transformations}


Before going further and discussing the transformation properties of these twisted fields under $SU(N)$ group and construction of various Hamiltonian densities, we would
like to discuss a very convenient map between the twisted creation/annihilation operators and their untwisted counterparts.  Such a map not only 
enables us to do various cumbersome manipulations on twisted operators in a convenient way but will also enable us to compare and contrast the twisted theories with their untwisted
counterparts. 

We start with noting the fact that we can define certain composite operators as
\begin{eqnarray}
 \tilde{a}_r & = & c_r \, e^{-\frac{i}{2} \lambda^{(r)} \wedge Q}, \nonumber \\
\tilde{a}^\dagger_r & = &  e^{\frac{i}{2} \lambda^{(r)} \wedge Q} \, c^\dagger_r, \nonumber \\
 \tilde{b}_r & = & d_r \, e^{\frac{i}{2} \lambda^{(r)} \wedge Q}, \nonumber \\
\tilde{b}^\dagger_r & = &  e^{-\frac{i}{2} \lambda^{(r)} \wedge Q} \, d^\dagger_r, 
\label{dresstranformagen}
\end{eqnarray}
where $Q_m$; $m = 1, 2, \cdots, (N-1)$ are the Cartans of the $SU(N)$ group, given by (\ref{fockcargensun}) and the $\lambda^{(r)}_m$ are defined by 
(\ref{weightcre}), (\ref{weightani}). 
Also, $\lambda^{(r)} \wedge Q = \lambda^{(r)}_l \theta_{lm} Q_m $; $l, m = 1, 2, \cdots (N - 1)$, $\theta_{lm} = - \theta_{ml}$ being an arbitrary real anti-symmetric matrix.

One can check that operators in (\ref{dresstranformagen}) satisfy the same twisted commutation relations as (\ref{atwistedcomrel}) if we identify 
$\lambda^{(r)}_l \theta_{lm} \lambda^{(s)}_m = \tilde{\lambda}^{(r)}_{l} \tilde{\theta}_{lm} \tilde{\lambda}^{(s)}_{m}$. But as both $\tilde{\theta}_{lm}$ and $\theta_{lm}$ are
arbitrary matrices, the above demand is always satisfied. 

Hence we have a map between creation/annihilation operators of the twisted theory with those of the untwisted theory, which we call as ``dressing transformations'' and is given by
\begin{eqnarray}
 a_r \, \, = &   c_r \, e^{-\frac{i}{2} \lambda^{(r)} \wedge Q}   & = \,  \, e^{-\frac{i}{2} \lambda^{(r)} \wedge Q} \, c_r,  \nonumber \\
a^\dagger_r \, \, = &   c^\dagger_r \, e^{\frac{i}{2} \lambda^{(r)} \wedge Q}  & = \, \, e^{\frac{i}{2} \lambda^{(r)} \wedge Q} \, c^\dagger_r ,  \nonumber \\
 b_r  \,  \,= &  d_r \, e^{\frac{i}{2} \lambda^{(r)} \wedge Q}  & =  \, \, e^{\frac{i}{2} \lambda^{(r)} \wedge Q}  \, d_r, \nonumber \\
b^\dagger_r  \, \, = & d^\dagger_r  \,  e^{-\frac{i}{2} \lambda^{(r)} \wedge Q}  & = \, \, e^{-\frac{i}{2} \lambda^{(r)} \wedge Q} \, d^\dagger_r.
\label{dresstranformanti}
\end{eqnarray}
This dressing map extends to all operators and state vectors in the two theories and provides us with a convenient way to discuss the twisted field theories.  For twisted fields of 
(\ref{modeexpansionatwist}), we have 
\begin{eqnarray}
 \phi_{\theta, r} (x) \, \, = &  \phi_{0, r} (x) \, e^{-\frac{i}{2} \lambda^{(r)} \wedge Q} & = \, \, e^{-\frac{i}{2} \lambda^{(r)} \wedge Q} \,  \phi_{0, r} (x),\nonumber \\
 \phi^\dagger_{\theta, r} (x) \, \, = &  \phi^\dagger_{0, r} (x) \, e^{\frac{i}{2} \lambda^{(r)} \wedge Q} & = \, \, e^{\frac{i}{2} \lambda^{(r)} \wedge Q} \, \phi^\dagger_{0, r} (x),
\label{fieldsantidress}
\end{eqnarray}
where $\phi_{0, r} (x)$, $\phi^\dagger_{0, r} (x)$ are the untwisted fields given by (\ref{modeexpansioncom}) and we have put a subscript ``0'' to distinguish them from twisted fields.
Also we note that $\lambda^{(r)} \wedge \lambda^{(r)} = 0$, owing to the antisymmetry of $\theta$. Hence one can freely move the exponential terms in (\ref{dresstranformanti}) and
(\ref{fieldsantidress}) from left to right and vice versa.  The antisymmetry of the $\theta$ matrix also means that it is not possible to get twisted statistics for any 
internal symmetry group which is of rank less than 2. Thus $SU(3)$ is the smallest $SU(N)$ group for which we can have a twisted statistics of the above type.

The twisted fields satisfy the commutation relations
\begin{eqnarray}
 \phi_{\theta,r}(x) \phi_{\theta,s}(x) & = & e^{ i \lambda^{(r)} \wedge \lambda^{(s)} } \, \phi_{\theta,s}(x) \phi_{\theta,r}(x), \nonumber \\
\phi^{\dagger}_{\theta,r}(x) \phi^{\dagger}_{\theta,s}(x) & = & e^{ i \lambda^{(r)} \wedge \lambda^{(s)} } \, \phi^{\dagger}_{\theta,s}(x) \phi^{\dagger}_{\theta,r}(x),
\label{antitwistfield}
\end{eqnarray}
which can be easily checked by using (\ref{atwistedcomrel}) or alternatively by using (\ref{fieldsantidress}). 

We also note that, the number operator $N$ remains unchanged i.e.
\begin{eqnarray}
N_\theta & = &  \int \frac{d^3 p}{(2\pi)^3} \frac{1}{2E_{p}} \left[ a^\dagger_r a_r \, + \, b^\dagger_r b_r \right] \nonumber \\
& = &  \int \frac{d^3 p}{(2\pi)^3} \frac{1}{2E_{p}} \left[ c^\dagger_r c_r \, + \, d^\dagger_r d_r \right] \, = \, N_0.
\end{eqnarray}
Also we have 
\begin{eqnarray}
 \left[ Q_m , \phi_{\theta, r} (x) \right ] & = & -\lambda^{(r)}_m \phi_{\theta, r} (x),  \nonumber \\
 \left[ Q_m , \phi^{\dagger}_{\theta, r} (x) \right ] & = &  \lambda^{(r)}_m \phi^{\dagger}_{\theta, r} (x),
\label{relationcomso6}
\end{eqnarray}
which is same as the relation (\ref{chargefields}) satisfied by the untwisted fields. It should be noted that, in (\ref{relationcomso6}) the charge operators $ Q_m$ are
taken in the untwisted form, as only the untwisted charge operators furnish the correct Fock space representation of the given $SU(N)$ group. Furthermore, it can be easily seen that 
the ``charge operators'' constructed using twisted creation/annihilation operators do not satisfy the Lie algebra of the $SU(N)$ group and hence do not furnish a Fock space representation for the 
$SU(N)$ group.

Also, if $| 0 \rangle_0$ and $| 0 \rangle_\theta$  are the vacua of untwisted and twisted theories then we have
\begin{eqnarray}
c_r  | 0 \rangle_0 & = &  d_r  | 0 \rangle_0 \, = \, 0,  \nonumber \\
a_r  | 0 \rangle_\theta & = &  b_r  | 0 \rangle_\theta \, = \, 0 .
\label{vaccuas}
\end{eqnarray}
But because of the dressing transformations (\ref{dresstranformanti}) we have
\begin{eqnarray}
a_r  | 0 \rangle_0 & = &  c_r \, e^{-\frac{i}{2} \lambda^{(r)} \wedge Q} | 0 \rangle_0 ,\nonumber \\
b_r | 0 \rangle_0 & = &  d_r \, e^{\frac{i}{2} \lambda^{(r)} \wedge Q} | 0 \rangle_0.   
\label{vaccuasdress}
\end{eqnarray}
If the untwisted vacuum is invariant under the $SU(N)$ group transformations i.e. if $Q_m | 0 \rangle_0 = E_n | 0 \rangle_0 = 0$ then
\begin{eqnarray}
a_r  | 0 \rangle_0 & = &  c_r  | 0 \rangle_0 \, = \, 0, \nonumber \\
b_r | 0 \rangle_0 & = &  d_r | 0 \rangle_0   \, = \, 0.
\label{vaccuasinv}
\end{eqnarray}
Hence, the vacuum of the two theories is one and same. Similarly we find that, provided the untwisted vacuum is invariant under the $SU(N)$ group transformations, 
the single-particle states in the two theories are also same i.e.
\begin{eqnarray}
\overline{| r \rangle}_\theta & = &  \overline{|r \rangle}_0 , \nonumber \\
| r \rangle_\theta & = &  |r \rangle_0 .  
\label{singlrpart}
\end{eqnarray}
In this work we will restrict ourself only to the case of the untwisted vacuum being invariant under the $SU(N)$ group transformations. 
The case of it not being invariant under $SU(N)$ transformations is also a exciting scenario, as it will lift the degeneracy of the vacua of the two theories and 
we expect it to result into new features in the twisted theory. We plan to discuss 
it in more details in a separate work. 

Now we can discuss the transformation property of the fields $\phi_{\theta, r}$ under $SU(N)$, which is given by
\begin{eqnarray}
 U(\sigma) \phi_{\theta, r} (x) U^\dagger(\sigma) & = &  \phi'_{\theta, r} (x) \, = \,\, U(\sigma) e^{-\frac{i}{2} \lambda^{(r)} \wedge Q} U^\dagger(\sigma) \,
 U(\sigma) \phi_{0, r} (x) U^\dagger(\sigma) \nonumber \\ 
& = & U(\sigma) e^{-\frac{i}{2} \lambda^{(r)} \wedge Q} U^\dagger(\sigma) \, \left( e^{-i \sigma_a T_a} \right)_{rs} \,\phi_s (x) \, 
\, = \, \xi_{(r)}(\sigma) \, \left( e^{-i \sigma_a T_a} \right)_{rs} \,\phi_s (x),  \nonumber \\
 U(\sigma) \phi^\dagger_{\theta, r} (x) U^\dagger(\sigma) & = &  \phi'^\dagger_{\theta, r} (x) \, = \, U(\sigma) \phi^\dagger_{0, r} (x) U^\dagger(\sigma) 
\, U(\sigma) e^{\frac{i}{2} \lambda^{(r)} \wedge Q} U^\dagger(\sigma) \nonumber \\ 
& = & \left( e^{i \sigma_a T_a} \right)_{sr} \,\phi^\dagger_s (x) \, U(\sigma) e^{\frac{i}{2} \lambda^{(r)} \wedge Q} U^\dagger(\sigma) 
\, = \,\left( e^{i \sigma_a T_a} \right)_{sr} \,\phi^\dagger_s (x) \, \xi^\dagger_{(r)}(\sigma), 
\label{suntranformanti}
\end{eqnarray}
where $\xi_{(r)}(\sigma) = U(\sigma) e^{-\frac{i}{2} \lambda^{(r)} \wedge Q} U^\dagger(\sigma)$ is a unitary operator satisfying 
$\xi_{(r)}(\sigma)\xi^\dagger_{(r)}(\sigma) =  \xi^\dagger_{(r)}(\sigma) \xi_{(r)}(\sigma) = \textbf{I}$. 

Also we have 
\begin{eqnarray}
 \xi_{(r)}(\sigma) \, \left( e^{-i \sigma_a T_a} \right)_{rs} \,\phi_s (x) & = & U(\sigma) e^{-\frac{i}{2} \lambda^{(r)} \wedge Q} U^\dagger(\sigma) \,
 U(\sigma) \phi_{0, r} (x) U^\dagger(\sigma) \nonumber \\ 
& = & U(\sigma) e^{-\frac{i}{2} \lambda^{(r)} \wedge Q} \phi_{0, r} (x) U^\dagger(\sigma) \, = \, U(\sigma) \phi_{0, r} (x) e^{-\frac{i}{2} \lambda^{(r)} \wedge Q} U^\dagger(\sigma)
\nonumber \\
& = & U(\sigma) \phi_{0, r} (x) U^\dagger(\sigma) \, U(\sigma) e^{-\frac{i}{2} \lambda^{(r)} \wedge Q} U^\dagger(\sigma) 
\, = \, \left( e^{-i \sigma_a T_a} \right)_{rs} \,\phi_s (x) \, \xi_{(r)}(\sigma).
\label{antixifree}
\end{eqnarray}
The transformation properties of the state vectors can be similarly discussed. For example, assuming that the vacuum remains invariant under the $SU(N)$ transformations 
i.e. $ U(\sigma) | 0 \rangle = | 0 \rangle$, the single-particle states transform as
\begin{eqnarray}
 U(\sigma) | r \rangle_\theta & = &  U(\sigma) a^\dagger_r | 0 \rangle \, = \, U(\sigma) a^\dagger_r U^\dagger(\sigma) U(\sigma) | 0 \rangle 
 \, = \, U(\sigma) c^\dagger_r \, e^{\frac{i}{2} \lambda^{(r)} \wedge Q} U^\dagger(\sigma) U(\sigma) | 0 \rangle \nonumber \\ 
& = &  U(\sigma) c^\dagger_r U^\dagger(\sigma)| 0 \rangle \, = \, \left( e^{i \sigma_a T_a} \right)_{sr} \,c^\dagger_s | 0 \rangle 
\, = \, \left( e^{i \sigma_a T_a} \right)_{sr} \,a^\dagger_s | 0 \rangle \, = \, \left( e^{i \sigma_a T_a} \right)_{sr}  | s \rangle_\theta, \nonumber \\
 U(\sigma) \overline{| r \rangle}_\theta & = &  U(\sigma) b^\dagger_r | 0 \rangle \, = \, U(\sigma) b^\dagger_r U^\dagger(\sigma) U(\sigma) | 0 \rangle 
 \, = \, U(\sigma) d^\dagger_r \, e^{-\frac{i}{2} \lambda^{(r)} \wedge Q} U^\dagger(\sigma) U(\sigma) | 0 \rangle \nonumber \\
& = & U(\sigma) d^\dagger_r U^\dagger(\sigma)| 0 \rangle \, = \, \left( e^{-i \sigma_a T^\ast_a} \right)_{sr} \,d^\dagger_s | 0 \rangle
\, = \, \left( e^{-i \sigma_a T^\ast_a} \right)_{sr} \,b^\dagger_s | 0 \rangle  \, = \, \left( e^{-i \sigma_a T^\ast_a} \right)_{sr}  \overline{| s \rangle}_\theta.
\label{1tstatevecsuntransform}
\end{eqnarray}
Hence, under $SU(N)$ transformations, the twisted single-particle states transform in same way as the untwisted single-particle states. Due to (\ref{singlrpart}), the transformation 
of twisted single-particle states was expected to be same as that of the untwisted state. However, twisted multi-particle states will have a different transformation 
e.g. for two-particle state we have 
\begin{eqnarray}
 U(\sigma) | r , s \rangle_\theta & = &  U(\sigma) a^\dagger_s a^\dagger_r | 0 \rangle 
\, = \,  U(\sigma) a^\dagger_s U^\dagger(\sigma) \, U(\sigma) a^\dagger_r U^\dagger(\sigma) \, U(\sigma) | 0 \rangle \nonumber \\ 
& = & U(\sigma) c^\dagger_s \, e^{\frac{i}{2} \lambda^{(s)} \wedge Q} U^\dagger(\sigma) \, 
U(\sigma) c^\dagger_r \, e^{\frac{i}{2} \lambda^{(r)} \wedge Q} U^\dagger(\sigma) \, U(\sigma) | 0 \rangle \nonumber \\ 
& = &   e^{\frac{i}{2} \lambda^{(s)} \wedge \lambda^{(r)}} \, U(\sigma) c^\dagger_s U^\dagger(\sigma) \, U(\sigma) c^\dagger_r U^\dagger(\sigma)| 0 \rangle \nonumber \\ 
& = &  e^{\frac{i}{2} \lambda^{(s)} \wedge \lambda^{(r)}} \, \left( e^{i \sigma_a T_a} \right)_{ts} \left( e^{i \sigma_a T_a} \right)_{ur} \, c^\dagger_t c^\dagger_u | 0 \rangle 
 \nonumber \\ 
& = & e^{\frac{i}{2} \lambda^{(s)} \wedge \lambda^{(r)}} \, e^{-\frac{i}{2} \lambda^{(t)} \wedge \lambda^{(u)}} \, \left( e^{i \sigma_a T_a} \right)_{ts}  
\left( e^{i \sigma_a T_a} \right)_{ur} \, a^\dagger_t a^\dagger_u| 0 \rangle \nonumber \\ 
& = &  e^{\frac{i}{2} \lambda^{(s)} \wedge \lambda^{(r)}} \, e^{-\frac{i}{2} \lambda^{(t)} \wedge \lambda^{(u)}} \, \left( e^{i \sigma_a T_a} \right)_{ts}  
\left( e^{i \sigma_a T_a} \right)_{ur} \, | u ,t \rangle_\theta ,
\label{2atstatevecsuntransform}
\end{eqnarray}
and 
\begin{eqnarray}
 U(\sigma) \overline{| r, s \rangle}_\theta & = & U(\sigma) b^\dagger_s b^\dagger_r | 0 \rangle 
\, = \,  U(\sigma) b^\dagger_s U^\dagger(\sigma) \, U(\sigma) b^\dagger_r U^\dagger(\sigma) \, U(\sigma) | 0 \rangle \nonumber \\ 
& = & U(\sigma) d^\dagger_s \, e^{-\frac{i}{2} \lambda^{(s)} \wedge Q} U^\dagger(\sigma) \, 
U(\sigma) d^\dagger_r \, e^{-\frac{i}{2} \lambda^{(r)} \wedge Q} U^\dagger(\sigma) \, U(\sigma) | 0 \rangle\nonumber \\
& = &  e^{\frac{i}{2} \lambda^{(s)} \wedge \lambda^{(r)}} \, U(\sigma) d^\dagger_s U^\dagger(\sigma) \, U(\sigma) d^\dagger_r U^\dagger(\sigma)| 0 \rangle \nonumber \\ 
& = & e^{\frac{i}{2} \lambda^{(s)} \wedge \lambda^{(r)}}\, \left( e^{-i \sigma_a T^\ast_a} \right)_{ts}\, \left( e^{-i \sigma_a T^\ast_a} \right)_{ur}\,d^\dagger_t d^\dagger_u| 0 \rangle
\nonumber \\ 
& = & e^{\frac{i}{2} \lambda^{(s)} \wedge \lambda^{(r)}}\, e^{-\frac{i}{2} \lambda^{(t)} \wedge \lambda^{(u)}}\, \left( e^{-i \sigma_a T^\ast_a} \right)_{ts}\, 
\left( e^{-i \sigma_a T^\ast_a} \right)_{ur}\, b^\dagger_t b^\dagger_u| 0 \rangle \nonumber \\ 
& = &  e^{\frac{i}{2} \lambda^{(s)} \wedge \lambda^{(r)}}\, e^{-\frac{i}{2} \lambda^{(t)} \wedge \lambda^{(u)}}\, \left( e^{-i \sigma_a T^\ast_a} \right)_{ts}\, 
\left( e^{-i \sigma_a T^\ast_a} \right)_{ur}\, \overline{| u, t \rangle}_\theta.
\label{2btstatevecsuntransform}
\end{eqnarray}
Similarly all other multi-particle states follow twisted transformation rules.

Before we write down field theories using $\phi_{\theta, r}$ fields, we define a new multiplication rule for the product of two fields. We define the `` antisymmetric star-product '' 
$\star$ as follows
\begin{eqnarray}
 \phi^\#_{\theta, r} (x) \star \phi^\#_{\theta, s}(y) & = &  \phi^\#_{\theta, r} (x) \, e^{-\frac{i}{2} \overleftarrow{Q} \wedge \overrightarrow{Q} } \phi^\#_{\theta, s} (y)\nonumber \\
& = &  \phi^\#_{\theta, r} (x) \phi^\#_{\theta, s} (y) \, - \, \frac{i}{2} \theta_{lm} \left[ Q_l, \phi^\#_{\theta, r} (x)\right] \left[ Q_m, \phi^\#_{\theta, s} (y) \right]\nonumber \\
& + & \frac{1}{2!} \frac{i}{2} \theta_{lm} \, \frac{i}{2} \theta_{np} \left[ Q_l, \left[ Q_n,\phi^\#_{\theta, r} (x)\right]\right] 
\left[ Q_m, \left[ Q_p,\phi^\#_{\theta, s} (y) \right] \right]
\, + \, \cdots,
\label{starproductanti}
\end{eqnarray}
where $\phi^\#_{\theta, r}$ stands for either of $\phi_{\theta, r}$ or $\phi^\dagger_{\theta, r}$. It should be remarked that the above defined $\star$-product is infact the 
internal symmetry analogue of the widely studied Moyal-product \cite{bal-pinzul}. 

Note that, owing to the relation (\ref{relationcomso6}), we have 
\begin{eqnarray}
 \phi^\#_{\theta, r} (x) \star \phi^\#_{\theta, s}(y) & = & e^{-\frac{i}{2} (\pm \lambda^{(r)}) \wedge (\pm \lambda^{(s)}) } \phi^\#_{\theta, r} (x) 
\, \cdot \, \phi^\#_{\theta, s}(y),
\label{starproductantirel}
\end{eqnarray}
where $\lambda^{(r)}$ should be taken if $\phi^\#_{\theta, r} = \phi^\dagger_{\theta, r}$ and $-\lambda^{(r)}$ should be taken if $\phi^\#_{\theta, r} = \phi_{\theta, r}$.
Hence, multiplying two fields with a star-product is nothing but multiplying a certain phase factor to ordinary product of fields. Nonetheless, using the star-product 
greatly simplifies things and it should be regarded just as a shorthand notation for the phases that are present in a particular term in the Hamiltonian density. These 
simplifications will be further elaborated when we construct the interaction terms for twisted fields.

The star-product introduced here has the property that
\begin{eqnarray}
 \phi^\#_{\theta,r} \, \star \, \phi^\#_{\theta, s} & = &  \phi^\#_{\theta, s} \, \star \, \phi^\#_{\theta,r}, \nonumber \\
 \phi^\#_{\theta,r} \, \star \, \left( \phi^\#_{\theta, s} \, \star \, \phi^\#_{\theta, t} \right) 
& = & \left( \phi^\#_{\theta,r} \, \star \, \phi^\#_{\theta, s} \right)\, \star \, \phi^\#_{\theta, t}, 
\label{idanti1}
\end{eqnarray}
and due to the antisymmetry we have 
\begin{eqnarray}
 \phi^\#_{\theta,r} \, \star \, \phi^\#_{\theta, r} & = &  \phi^\#_{\theta, r} \cdot \phi^\#_{\theta,r}. 
\label{idanti2}
\end{eqnarray}
Also, because of the dressing transformations (\ref{fieldsantidress}) we have
\begin{eqnarray}
 \phi^\#_{\theta,r_1} \, \star \, \phi^\#_{\theta, r_2} \, \star \, \cdots \, \star \, \phi^\#_{\theta, r_n} & = & \phi^\#_{0,r_1} \, \phi^\#_{0, r_2} \, \cdots \, \phi^\#_{0, r_n} 
\, e^{\frac{i}{2} \left( \pm \lambda^{(r_1)} \, \pm \,  \lambda^{(r_2)} \, \pm \, \cdots \, \pm \,  \lambda^{(r_n)} \right) \wedge Q },
\label{idanti3}
\end{eqnarray}
where $ + \lambda^{(r)}$ is to be taken if $ \phi^\#_{\theta,r} = \phi^\dagger_{\theta,r}$ and $ - \lambda^{(r)}$ if $ \phi^\#_{\theta,r} = \phi_{\theta,r}$.

Having introduced the $\star$-product we can now discuss how to write Hamiltonian densities using twisted fields. For any given untwisted Hamiltonian its twisted counterpart 
should be written by replacing the untwisted fields $\phi_{0,r}$ by the twisted fields $\phi_{\theta,r}$ and the ordinary product between fields by the $\star$-product. 
Hence the free theory Hamiltonian density $\mathcal{H}_{\theta, F}$ in terms of twisted fields becomes
 \begin{eqnarray}
  \mathcal{H}_{\theta, F} & = &  \Pi^{\dagger}_{\theta,r} \star \Pi_{\theta,r}  \, + \, (\partial_i \phi^{\dagger}_{\theta,r}) \star(\partial^i \phi_{\theta,r})
 \, + \, m^2  \, \phi^{\dagger}_{\theta,r} \star \phi_{\theta,r} \nonumber \\
& = &  \Pi^{\dagger}_{\theta,r} \Pi_{\theta,r}  \, + \, (\partial_i \phi^{\dagger}_{\theta,r}) (\partial^i \phi_{\theta,r})  \, + \, m^2  \, \phi^{\dagger}_{\theta,r} \phi_{\theta,r},
\label{freeantiham}
 \end{eqnarray}
where $\Pi_{\theta,r}$ is the canonical conjugate of $\phi_{\theta,r}$; $r = 1, 2, \cdots, N$ and the last line in (\ref{freeantiham}) is obtained using (\ref{idanti2}). 
This Hamiltonian density is invariant under the $SU(N)$ global transformations which can be explicitly checked by using (\ref{suntranformanti}). 

The renormalizable $SU(N)$ invariant interaction Hamiltonian density compatible with (\ref{hamsun}) is given by
\begin{eqnarray}
 \mathcal{H}_{\theta,\rm{Int}} & = &  \frac{\gamma}{4} \, \phi^{\dagger}_{\theta,r}  \star  \phi^{\dagger}_{\theta,s}  \star \phi_{\theta,r}  \star \phi_{\theta,s} \nonumber \\
& = & \frac{\gamma}{4}\, e^{- i \lambda^{(r)} \wedge \lambda^{(s)}} \, \phi^{\dagger}_{\theta,r} \phi^{\dagger}_{\theta,s} \phi_{\theta,r} \phi_{\theta,s},
\label{intantiham}
\end{eqnarray}
where $r,s = 1, 2, \cdots, N$. 

Using (\ref{suntranformanti}) one can check that (\ref{intantiham}) is indeed invariant under $SU(N)$ group. Alternatively, one can also use the dressing transformation 
(\ref{fieldsantidress}) and the identity (\ref{idanti2}) to write everything in terms of the untwisted fields and then apply the $SU(N)$ transformations given by 
(\ref{suntranformcom}) to check for the invariance of the Hamiltonian density. 

From (\ref{intantiham}) it is clear that unlike the untwisted case where $SU(N)$ invariance forces all the interaction terms in (\ref{gaugen4susy}) to have the same coupling $\gamma$, 
here the demand of $SU(N)$ invariance forces the various terms to have different couplings related with each other by phases of the type $e^{\pm i \lambda^{(r)} \wedge \lambda^{(s)}}$.

Although in this section we restricted our discussion only to scalar fields and twisted bosons but it is easy to generalize the discussion to include twisted fermions and 
spinor fields. For the discussion of twisted fermions we have to 
consider anticommuting creation/annihilation operators. The twisted fermions will again satisfy (\ref{atwistedcomrel}) but with $\eta = -1$. One can again write down $SU(N)$
invariant field theories involving such twisted fermions in a way very similar to the one we discussed.  Also, the above discussion can be easily 
generalized to higher dimensional representations of $SU(N)$ group as well as to other symmetry groups like $SO(N)$.


\subsection{$S$-matrix Elements}


In the previous section we set up the formalism for writing down field theories with a special type of twisted statistics which we called `` antisymmetric twisted statistics''.
We showed that using such twisted fields we can write down $SU(N)$ invariant interactions. We now want to discuss the possibility of experimental signatures of the twisted 
field theories. We start our discussion from scattering processes. 

Let us first start with the $SU(N)$ invariant free Hamiltonian density $\mathcal{H}_{\theta, F}$. Using the dressing transformation (\ref{fieldsantidress}) and the 
identity (\ref{idanti3}) we have
 \begin{eqnarray}
  \mathcal{H}_{\theta, F} & = &  \Pi^{\dagger}_{\theta,r} \star \Pi_{\theta,r}  \, + \, (\partial_i \phi^{\dagger}_{\theta,r}) \star(\partial^i \phi_{\theta,r})
 \, + \, m^2  \, \phi^{\dagger}_{\theta,r} \star \phi_{\theta,r} \nonumber \\
& = &  \Pi^{\dagger}_{0,r} \Pi_{0,r}  \, + \, (\partial_i \phi^{\dagger}_{0,r}) (\partial^i \phi_{0,r}) \, + \, m^2  \, \phi^{\dagger}_{0,r} \phi_{0,r}
\, = \,  \mathcal{H}_{0,F} .
\label{freeantihamrel}
 \end{eqnarray}
Hence the Hamiltonian for twisted free theory is same as its untwisted counterpart. 
Next we look at the renormalizable $SU(N)$ invariant interacting Hamiltonian density which is
 \begin{eqnarray}
  \mathcal{H}_{\theta, \rm{I}} & = &  \Pi^{\dagger}_{\theta,r} \star \Pi_{\theta,r}  \, + \, (\partial_i \phi^{\dagger}_{\theta,r}) \star(\partial^i \phi_{\theta,r})
 \, + \, m^2  \, \phi^{\dagger}_{\theta,r} \star \phi_{\theta,r} 
\, + \,  \frac{\gamma}{4} \, \phi^{\dagger}_{\theta,r}  \star  \phi^{\dagger}_{\theta,s}  \star \phi_{\theta,r}  \star \phi_{\theta,s} \nonumber \\
& = &  \Pi^{\dagger}_{0,r} \Pi_{0,r}  \, + \, (\partial_i \phi^{\dagger}_{0,r}) (\partial^i \phi_{0,r}) \, + \, m^2  \, \phi^{\dagger}_{0,r} \phi_{0,r}
\, + \,  \frac{\gamma}{4} \, \phi^{\dagger}_{0,r}  \phi^{\dagger}_{0,s}  \phi_{0,r}  \phi_{0,s} 
 \, = \, \mathcal{H}_{0, \rm{I}} ,
\label{intantihamrel}
 \end{eqnarray}
where to obtain the last line we have again used the dressing transformation (\ref{fieldsantidress}) and the identity (\ref{idanti3}). So even the $SU(N)$ invariant interacting 
Hamiltonian density for the two theories turns out to be same. But the in/out states for the twisted theory also contain information about twisted statistics. 
So we should look at $S$-matrix elements which can still provide information about twisted statistics. Let us take a typical $S$-matrix element, say for the scattering process of 
$\phi_{\theta, r} \phi_{\theta, s} \rightarrow \phi_{\theta, r} \phi_{\theta, s}$. Then we have 
 \begin{eqnarray}
  S\left[ \phi_{\theta, r} \phi_{\theta, s} \rightarrow \phi_{\theta, r} \phi_{\theta, s}  \right] 
& = & \leftidx{_{out,\theta}}{\left \langle r s | r s  \right \rangle }{_{ \theta,in}} 
\, = \, \leftidx{_{\theta}}{\left \langle r s | S_\theta | r s \right \rangle }{_{ \theta}},
\label{smatanti}
 \end{eqnarray}
where $S_\theta = \mathcal{T}  \exp \left[-i\int^{\infty}_{-\infty} d^{4}z \mathcal{H}_{\theta, \rm{ Int}} (z) \right] $ is the $S$-operator and we have denoted the two-particle in and 
out states by $\left| r s \right \rangle_\theta = a^\dagger_s a^\dagger_r | 0 \rangle$. Because of (\ref{intantihamrel}) we have 
\begin{eqnarray}
 S_\theta & = & \mathcal{T}  \exp \left[-i\int^{\infty}_{-\infty} d^{4}z \mathcal{H}_{\theta, \rm{Int}} (z) \right]  
\, = \, \mathcal{T}  \exp \left[-i\int^{\infty}_{-\infty} d^{4}z \mathcal{H}_{0, \rm{Int}} (z) \right]  \, = \, S_0.
\label{sopanti}
\end{eqnarray}
Also we have 
\begin{eqnarray}
 \left| r s \right \rangle_\theta & = & a^\dagger_s a^\dagger_r  \left| 0 \right\rangle 
\, = \, c^\dagger_s \, e^{\frac{i}{2} \lambda^{(s)} \wedge Q} \, c^\dagger_r \, e^{\frac{i}{2} \lambda^{(r)} \wedge Q} \left| 0 \right\rangle 
\, = \, e^{\frac{i}{2}\lambda^{(s)} \wedge \lambda^{(r)} } \,  c^\dagger_s \, c^\dagger_r \, e^{\frac{i}{2} ( \lambda^{(r)} + \lambda^{(s)}) \wedge Q} \left| 0 \right\rangle 
 \nonumber \\
& = & e^{\frac{i}{2}\lambda^{(s)} \wedge \lambda^{(r)} } \,  c^\dagger_s \, c^\dagger_r \,\left| 0 \right\rangle  
\, = \, e^{\frac{i}{2}\lambda^{(s)} \wedge \lambda^{(r)} } \, \left| r s \right \rangle_0.
\label{stateanti}
\end{eqnarray}
Using (\ref{stateanti}) and (\ref{sopanti}) we get
 \begin{eqnarray}
  S\left[ \phi_{\theta, r} \phi_{\theta, s} \rightarrow \phi_{\theta, r} \phi_{\theta, s}  \right] & = & 
\leftidx{_{0}}{\left \langle r s | \, e^{\frac{i}{2}\lambda^{(s)} \wedge \lambda^{(r)} } \, S_0 \, e^{\frac{-i}{2}\lambda^{(s)} \wedge \lambda^{(r)} } \, | r s \right \rangle }{_{ 0}}
\nonumber \\
& = & \leftidx{_{0}}{\left \langle r s | \, S_0 \, | r s \right \rangle }{_{ 0}} \nonumber \\
& = &  S\left[ \phi_{0, r} \phi_{0, s} \rightarrow \phi_{0, r} \phi_{0, s}  \right]. 
\label{smatantirel}
\end{eqnarray}
So it turns out that, even the $S$-matrix elements for the two theories are same. It seems that the twisted $SU(N)$ invariant theory is indistinguishable from an untwisted 
$SU(N)$ invariant
 theory but we should take note of the fact that the particles in the two theory follow different type of statistics. Hence, the best place to look for potential signatures
of such particles is to look at the statistical properties and to construct observables which depend crucially on the statistics followed by these particles.  
 
One should also note that this indistinguishability arised because we demanded that our Hamiltonian density remains invariant under $SU(N)$ transformations and that 
the twisted vacuum is not only same as untwisted vacuum but also annihilates all the charge operators. Dropping either of these two demands makes the two theories distinct. 
We now briefly discuss the first scenario i.e. the case where the Hamiltonian density does not remain invariant under $SU(N)$ transformations. We plan to consider the 
second scenario in more details in a separate work.   

Let us take the Hamiltonian density (\ref{intantihamrel}) but with fields multiplied not by $\star$-products but by ordinary products i.e.
\begin{eqnarray}
  \mathcal{H}_{\theta} & = &  \Pi^{\dagger}_{\theta,r} \Pi_{\theta,r}  \, + \, (\partial_i \phi^{\dagger}_{\theta,r}) (\partial^i \phi_{\theta,r})
 \, + \, m^2  \, \phi^{\dagger}_{\theta,r} \phi_{\theta,r} 
\, + \,  \frac{\gamma}{4} \, \phi^{\dagger}_{\theta,r}  \phi^{\dagger}_{\theta,s} \phi_{\theta,r}  \phi_{\theta,s} \, + \, \rm{h. c}.
\label{hamwithoutstar}
\end{eqnarray}
Since $\phi_{\theta, r}$ are noncommuting fields so there is an ambiguity in ordering of operators in the interaction term and there are other inequivalent terms 
which one can write. The full Hamiltonian with all $\phi^4$ type terms can be written as 
 \begin{eqnarray}
  \mathcal{H}_{\theta} & = &  \Pi^{\dagger}_{\theta,r} \Pi_{\theta,r}  \, + \, (\partial_i \phi^{\dagger}_{\theta,r}) (\partial^i \phi_{\theta,r})
 \, + \, m^2  \, \phi^{\dagger}_{\theta,r} \phi_{\theta,r} 
\, + \,  \frac{\gamma}{4} \, \left[ \phi^{\dagger}_{\theta,r}  \phi^{\dagger}_{\theta,s} \phi_{\theta,r}  \phi_{\theta,s} 
\, + \, \phi^{\dagger}_{\theta,s} \phi^{\dagger}_{\theta,r} \phi_{\theta,r}  \phi_{\theta,s} 
\right . \nonumber \\  
& + & \left.\phi^{\dagger}_{\theta,s} \phi^{\dagger}_{\theta,r} \phi_{\theta,s} \phi_{\theta,r}
 \, + \, \phi^{\dagger}_{\theta,r}  \phi^{\dagger}_{\theta,s} \phi_{\theta,s} \phi_{\theta,r}  \, + \, \cdots  \right ] \, + \, \rm{h. c}.
\label{genhamwithoutstar}
\end{eqnarray}
One can in principle write 24 such terms and $\cdots$ represents the other terms which we have not written. Some of these 24 terms will be equivalent to other terms but 
unlike the untwisted case not all of them are equal to each other. Moreover, this Hamiltonian has no $SU(N)$ symmetry. The easiest way to see that is by using the 
dressing transformation (\ref{fieldsantidress}) and writing it in terms of untwisted fields 
 \begin{eqnarray}
\mathcal{H}_{\theta} & = &  \Pi^{\dagger}_{0,r} \Pi_{0,r}  \, + \, (\partial_i \phi^{\dagger}_{0,r}) (\partial^i \phi_{0,r})
 \, + \, m^2  \, \phi^{\dagger}_{0,r} \phi_{0,r}
\, + \,  \frac{\gamma}{4} \,e^{\frac{i}{2}\lambda^{(r)} \wedge \lambda^{(s)} } \,  \phi^{\dagger}_{0,r}  \phi^{\dagger}_{0,s} \phi_{0,r}  \phi_{0,s} 
\, + \, \frac{\gamma}{4} \, \phi^{\dagger}_{0,s} \phi^{\dagger}_{0,r} \phi_{0,r}  \phi_{0,s} \nonumber \\  
& + & \frac{\gamma}{4} \,e^{-\frac{i}{2}\lambda^{(r)} \wedge \lambda^{(s)} } \,  \phi^{\dagger}_{0,s} \phi^{\dagger}_{0,r}   \phi_{0,s} \phi_{0,r}  
\, + \, \frac{\gamma}{4} \, \phi^{\dagger}_{0,r}  \phi^{\dagger}_{0,s}   \phi_{0,s} \phi_{0,r}  \, + \, \cdots   \, + \, \rm{h. c}.
\label{genhamrelwithoutstar}
\end{eqnarray}
Because of the presence of $e^{\pm\frac{i}{2}\lambda^{(r)} \wedge \lambda^{(s)} }$  type phases, clearly this Hamiltonian density has no symmetry. Infact, (\ref{genhamwithoutstar}) is equivalent 
to a marginally deformed $SU(N)$ Hamiltonian density. We will discuss more about such Hamiltonian densities in the next section.

Similarly one can consider many more Hamiltonians which explicitly break $SU(N)$ invariance e.g. let us consider the interaction Hamiltonian given by
\begin{eqnarray}
  \mathcal{H}_{\theta,\rm{Int}} & = &  \frac{\gamma}{4} \, \phi^{\dagger}_{\theta,r}  \star  \phi^{\dagger}_{\theta,r}  \star \phi^{\dagger}_{\theta,s}  \star \phi_{\theta,s} 
\, + \, \rm{h.c}.
\label{hamsunvioanti}
\end{eqnarray}
Unlike (\ref{genhamwithoutstar}) whose untwisted counterpart was $SU(N)$ invariant, even the untwisted counterpart of this Hamiltonian 
\begin{eqnarray}
 \mathcal{H}_{0,\rm{Int}} & = &  \frac{\gamma}{4} \, \phi^{\dagger}_{0,r}    \phi^{\dagger}_{0,r}  \phi^{\dagger}_{0,s}   \phi_{0,s} 
\, + \, \rm{h.c},
\label{hamsunviocom}
\end{eqnarray}
is not $SU(N)$ invariant. 

But now (\ref{hamsunvioanti}) is not even equivalent to any local untwisted Hamiltonian, as after using dressing transformation we have
\begin{eqnarray}
 \mathcal{H}_{\theta,\rm{Int}} & = &  \frac{\gamma}{4} \, \phi^{\dagger}_{0,r} \phi^{\dagger}_{0,r} \phi^{\dagger}_{0,s} \phi_{0,s} \, e^{ i \lambda^{(r)} \wedge Q }
\, + \, \rm{h.c},
\label{hamsunvioantirel}
\end{eqnarray}
which is nonlocal because of the presence of nonlocal operators $Q_m$. 

So we find that, if we demand the twisted Hamiltonians to be $SU(N)$ invariant then they turn out to indistinguishable from untwisted Hamiltonians. But if we relax the demand of $SU(N)$
invariance then the two Hamiltonians are not exactly same. Infact many of such $SU(N)$ breaking Hamiltonians like (\ref{hamsunvioanti}) turn out to be nonlocal and can't be mapped to any
local untwisted Hamiltonian. Even in the case of $SU(N)$ invariant Hamiltonians, one can possibly construct observables which show signatures of the underlying twisted statistics.


\subsection{The $\mathcal{N} = 4$ SUSY Hamiltonian and its Marginal ($\beta$-) Deformations}


As a specific example of equivalence between twisted interaction Hamiltonians of the type (\ref{genhamwithoutstar}) and untwisted marginally deformed $SU(N)$ Hamiltonians, 
let us look at the scalar matter sector of $\mathcal{N}  =  4$ supersymmetric (SUSY) Yang-Mills theory in four dimensions and its marginal deformations \cite{Leigh:1995ep}. 
Although it is not difficult to generalize the discussion to include the fermionic sector also but to illustrate our point it is sufficient to show the equivalence only for
the scalar sector. 
The scalar sector of the SUSY theory consists of six real scalars $\phi_{0,r}$; $r = 1, 2, \cdots 6$ having a $SO(6)$ global symmetry. The scalars transform as fundamental 
representation of $SO(6)$ group or equivalently as the 6-dimensional representation of the $SU(4)$ group \cite{Frolov:2005iq, Frolov:2005dj}. 
These six real scalars can also be combined to form 3 complex scalars $\Phi_{0,r}$; $r = 1,2,3$. When written in terms of the complex fields only the $SU(3)$ subgroup of 
the full symmetry group is apparent and the three complex fields transform as the fundamental representation of $SU(3)$ group. The interaction term is given by
\begin{eqnarray}
 \mathcal{H}_{0,\rm{int}} & = &  \frac{g^2}{2} f_{ijk} f^{i}_{lm}\,  \epsilon_{rst}\epsilon_{uvt} \, \Phi^{\dagger j}_{0,r} \Phi^{\dagger k}_{0,s} \Phi^l_{0,u} \Phi^m_{0,v}.
\label{gaugeneral4susy}
\end{eqnarray}
Since we are not concerned with the details of the gauge theory, we will ``switch off'' the gauge interaction. So without gauge interactions we have (denoting coupling constant by 
$\frac{\tilde{\gamma}'}{4}$)
\begin{eqnarray}
 \mathcal{H}_{0,\rm{int}} & = &  \frac{\tilde{\gamma}'}{4} \,  \epsilon_{rst}\epsilon_{uvt} \, \Phi^{\dagger}_{0,r} \Phi^{\dagger}_{0,s} \Phi_{0,u} \Phi_{0,v} \nonumber \\
& = & \frac{\tilde{\gamma}'}{4} \, ( \delta_{r u} \delta_{s v} \, - \, \delta_{r v} \delta_{s u} )\, \Phi^{\dagger}_{0,r} \Phi^{\dagger}_{0,s} \Phi_{0,u} \Phi_{0,v} \nonumber \\
& = &\frac{\tilde{\gamma}'}{4} \, \Phi^{\dagger}_{0,r} \Phi^{\dagger}_{0,s} \Phi_{0,r} \Phi_{0,s} 
\; - \; \frac{\tilde{\gamma}'}{4} \, \Phi^{\dagger}_{0,r} \Phi^{\dagger}_{0,s} \Phi_{0,s} \Phi_{0,r}.
\label{gauge4susy}
\end{eqnarray}
Without the gauge interactions, the fields commute i.e. $ [\Phi_{0,r}\, , \, \Phi_{0,s}] \; = \; [\Phi^{\dagger}_{0,r}\, , \, \Phi^{\dagger}_{0,s}] \; = \; 0 $. 
Hence, in the untwisted case the interaction Hamiltonian vanishes \cite{Sohnius:1985qm}. Let us see what happens if we replace the untwisted fields by twisted fields in 
(\ref{gauge4susy}).\footnote{As discussed before, there is an ambiguity in writing the interaction Hamiltonian density with twisted fields. It turns out that the most general 
Hamiltonian density also leads to the same results as this one. Hence, for sake of simplicity we choose to work with this Hamiltonian density which captures all the essential points.}
In that case we have
\begin{eqnarray}
 \mathcal{H}_{\theta,\rm{int}} & = & \frac{\tilde{\gamma}'}{4} \, \Phi^{\dagger}_{\theta,r} \Phi^{\dagger}_{\theta,s} \Phi_{\theta,r} \Phi_{\theta,s} 
\; - \; \frac{\tilde{\gamma}'}{4}\, \Phi^{\dagger}_{\theta,r} \Phi^{\dagger}_{\theta,s} \Phi_{\theta,s} \Phi_{\theta,r}.
\label{tgaugen4susy}
\end{eqnarray}
Using the relations
 \begin{eqnarray}
 \Phi_{\theta,r}(x) \Phi_{\theta,s}(x) & = & e^{ i \lambda^{(r)} \wedge \lambda^{(s)} } \, \Phi_{\theta,s}(x) \Phi_{\theta,r}(x), \nonumber \\
\Phi^{\dagger}_{\theta,r}(x) \Phi^{\dagger}_{\theta,s}(x) & = & e^{ i \lambda^{(r)} \wedge \lambda^{(s)} } \, \Phi^{\dagger}_{\theta,s}(x) \Phi^{\dagger}_{\theta,r}(x), 
\label{so6flipfield}
\end{eqnarray}
we have 
\begin{eqnarray}
 \mathcal{H}_{\theta,\rm{int}} & = & \frac{\tilde{\gamma}'}{4} \, \Phi^{\dagger}_{\theta,r} \Phi^{\dagger}_{\theta,s} \Phi_{\theta,r} \Phi_{\theta,s}
 \; - \; \frac{\tilde{\gamma}'}{4} \,e^{-i \lambda^{(r)}\wedge \lambda^{(s)}} \, \Phi^{\dagger}_{\theta,r} \Phi^{\dagger}_{\theta,s} \Phi_{\theta,r} \Phi_{\theta,s} \nonumber \\
& = & \frac{\tilde{\gamma}'}{4} \,\left( 1\; - \; e^{-i \lambda^{(r)}\wedge \lambda^{(s)}} \right)\, \Phi^{\dagger}_{\theta,r} \Phi^{\dagger}_{\theta,s} \Phi_{\theta,r} 
\Phi_{\theta,s} .
\label{tgauge4susy}
\end{eqnarray}
So the twisted interaction Hamiltonian density does not vanish. We can use the dressing transformations between twisted and untwisted fields 
to write it in terms of untwisted fields only. Then we have   
\begin{eqnarray}
 \mathcal{H}_{\theta,\rm{int}} & = & \frac{\tilde{\gamma}'}{4} \, \left( 1\; - \; e^{-i \lambda^{(r)}\wedge \lambda^{(s)}} \right)\, \Phi^{\dagger}_{0,r} \, 
e^{\frac{i}{2}\lambda^{(r)}\wedge Q}\, 
\Phi^{\dagger}_{0,s} \, e^{\frac{i}{2}\lambda^{(s)}\wedge Q}\, \Phi_{0,r} \, e^{\frac{-i}{2}\lambda^{(r)}\wedge Q}\, \Phi_{0,s} \, e^{\frac{-i}{2}\lambda^{(s)}\wedge Q}\, \nonumber \\
& = & \frac{\tilde{\gamma}'}{4} \, \left( 1\; - \; e^{-i \lambda^{(r)}\wedge \lambda^{(s)}} \right)\,e^{i \lambda^{(r)}\wedge \lambda^{(s)}}\, \Phi^{\dagger}_{0,r} 
\Phi^{\dagger}_{0,s} \Phi_{0,r} \Phi_{0,s} \nonumber \\
& = & \frac{\tilde{\gamma}'}{4} \, \left(e^{i \lambda^{(r)}\wedge \lambda^{(s)}} \; - \; 1 \right)\,\Phi^{\dagger}_{0,r} \Phi^{\dagger}_{0,s} \Phi_{0,r} \Phi_{0,s}.
\label{tcgauge4susy}
\end{eqnarray}
where r,s = 1,2,3. Expanding in terms of component fields we get
\begin{eqnarray}
 \mathcal{H}_{\theta,\rm{int}} & = & \frac{\tilde{\gamma}'}{4} \, \left[ \left(e^{i \lambda^{(1)}\wedge \lambda^{(1)}} \; - \; 1 \right)
\,\Phi^{\dagger}_{0,1} \Phi^{\dagger}_{0,1} \Phi_{0,1} \Phi_{0,1} 
\, + \, \left(e^{i \lambda^{(2)}\wedge \lambda^{(2)}} \; - \; 1 \right)\,\Phi^{\dagger}_{0,2} \Phi^{\dagger}_{0,2} \Phi_{0,2} \Phi_{0,2} 
\right. \nonumber \\
& + &  \left(e^{i \lambda^{(3)}\wedge \lambda^{(3)}} \; - \; 1 \right)\,\Phi^{\dagger}_{0,3} \Phi^{\dagger}_{0,3} \Phi_{0,3} \Phi_{0,3} 
\, + \,\left(e^{i \lambda^{(1)}\wedge \lambda^{(2)}} \; - \; 1 \right)\,\Phi^{\dagger}_{0,1} \Phi^{\dagger}_{0,2} \Phi_{0,1} \Phi_{0,2} \nonumber \\
& + &  \left(e^{i \lambda^{(1)}\wedge \lambda^{(3)}} \; - \; 1 \right)\,\Phi^{\dagger}_{0,1} \Phi^{\dagger}_{0,3} \Phi_{0,1} \Phi_{0,3}  
\, + \, \left(e^{i \lambda^{(2)}\wedge \lambda^{(3)}} \; - \; 1 \right)\,\Phi^{\dagger}_{0,2} \Phi^{\dagger}_{0,3} \Phi_{0,2} \Phi_{0,3} \nonumber \\
& + &   \left(e^{i \lambda^{(2)}\wedge \lambda^{(1)}} \; - \; 1 \right)\,\Phi^{\dagger}_{0,2} \Phi^{\dagger}_{0,1} \Phi_{0,2} \Phi_{0,1} 
\, + \,\left(e^{i \lambda^{(3)}\wedge \lambda^{(1)}} \; - \; 1 \right)\,\Phi^{\dagger}_{0,3} \Phi^{\dagger}_{0,1} \Phi_{0,3} \Phi_{0,1} \nonumber \\
& + & \left. \left(e^{i \lambda^{(3)}\wedge \lambda^{(2)}} \; - \; 1 \right)\,\Phi^{\dagger}_{0,3} \Phi^{\dagger}_{0,2} \Phi_{0,3} \Phi_{0,2}  \right].
\label{tcexpandgauge4susy}
\end{eqnarray}
Noting the fact that $e^{i \lambda^{(r)}\wedge \lambda^{(r)}} \, = \, 1$ and  $e^{i \lambda^{(s)}\wedge \lambda^{(r)}} \, = \, e^{-i \lambda^{(r)}\wedge \lambda^{(s)}} $, we 
can simplify (\ref{tcexpandgauge4susy}) and get 
\begin{eqnarray}
 \mathcal{H}_{\theta,\rm{int}} & = & \frac{\tilde{\gamma}'}{4} \, \left[ \left(e^{i \lambda^{(1)}\wedge \lambda^{(2)}} 
\; + \; e^{-i \lambda^{(1)}\wedge \lambda^{(2)}} \; - \;2 \right)\,\Phi^{\dagger}_{0,1} \Phi^{\dagger}_{0,2} \Phi_{0,1} \Phi_{0,2} 
\, + \, \left(e^{i \lambda^{(2)}\wedge \lambda^{(3)}} \; + \; e^{-i \lambda^{(2)}\wedge \lambda^{(3)}} \; - \;2 \right) \right. \nonumber \\
& & \left. \Phi^{\dagger}_{0,2} \Phi^{\dagger}_{0,3} \Phi_{0,2} \Phi_{0,3} 
\, + \, \left( e^{i \lambda^{(3)}\wedge \lambda^{(1)}} \; + \; e^{-i \lambda^{(3)}\wedge \lambda^{(1)}} \; - \;2 \right) \,
\Phi^{\dagger}_{0,3} \Phi^{\dagger}_{0,1} \Phi_{0,3} \Phi_{0,1} \right] \nonumber \\
& = & -\frac{\tilde{\gamma}'}{4} \, \left[ 4 \sin^2 \left\{ \frac{\lambda^{(1)}\wedge \lambda^{(2)}}{2} \right\}\, \Phi^{\dagger}_{0,1} \Phi^{\dagger}_{0,2} 
\Phi_{0,1} \Phi_{0,2} 
\, + \,  4 \sin^2 \left\{ \frac{\lambda^{(2)}\wedge \lambda^{(3)}}{2} \right\}\, \Phi^{\dagger}_{0,2} \Phi^{\dagger}_{0,3} \Phi_{0,2} \Phi_{0,3} \right. 
\nonumber \\
& + & \left. 4 \sin^2 \left\{ \frac{\lambda^{(3)}\wedge \lambda^{(1)}}{2} \right\}\,\Phi^{\dagger}_{0,3} \Phi^{\dagger}_{0,1} \Phi_{0,3} \Phi_{0,1} \right] 
\nonumber \\
& = &  \frac{\gamma_{12}}{2}\: \Phi^{\dagger}_{0,1} \Phi^{\dagger}_{0,2} \Phi_{0,1} \Phi_{0,2}  
\; + \; \frac{\gamma_{23}}{2} \: \Phi^{\dagger}_{0,2} \Phi^{\dagger}_{0,3} \Phi_{0,2} \Phi_{0,3}  
\; + \; \frac{\gamma_{31}}{2} \: \Phi^{\dagger}_{0,3} \Phi^{\dagger}_{0,1} \Phi_{0,3} \Phi_{0,1}.
\label{tcepgauge4susy}
\end{eqnarray}
We now show that (\ref{tcepgauge4susy}) is equivalent to marginal deformations of the scalar part of $\mathcal{N} = 4$ SUSY theory with gauge interactions switched off. 
The ``Marginally Deformed'' $\mathcal{N} = 4$ SUSY Hamiltonian density is given by \cite{Frolov:2005iq, Frolov:2005dj}
\begin{eqnarray}
\mathcal{H}_{0, \rm{int}} & = & \frac{\gamma}{4} \; Tr \left [ \left | \Phi_{0,1} \Phi_{0,2} - e^{-2i \pi \beta_{12}} \Phi_{0,2} \Phi_{0,1} \right |^2  
\; + \; \left | \Phi_{0,2} \Phi_{0,3} - e^{-2i \pi \beta_{23}}\Phi_{0,3} \Phi_{0,2} \right |^2  \right. \nonumber \\
& + & \left. \left | \Phi_{0,3} \Phi_{0,1} - e^{-2i \pi \beta_{31}} \Phi_{0,1} \Phi_{0,3} \right |^2 \right ] 
\; + \; \frac{\tilde{\gamma}}{4} \; Tr \left [ \left \{ [\Phi_{0,1} , \Phi^{\dagger}_{0,1}] + [\Phi_{0,2} , \Phi^{\dagger}_{0,2}] 
+ [\Phi_{0,3} , \Phi^{\dagger}_{0,3}] \right\}^2 \right ],
\label{mdgaugen4susy}
\end{eqnarray}
where the trace is over gauge index of the gauge group $SU(N)$. Since we are not interested in gauge fields, so we switch off the gauge interactions. 
The $ \mathcal{H}_{0, \rm{int}}$ then takes the form
\begin{eqnarray}
 \mathcal{H}_{0, \rm{int}} & = &  \frac{\gamma}{4} \left| \Phi_{0,1} \Phi_{0,2} - e^{-2i \pi \beta_{12}} \Phi_{0,2} \Phi_{0,1} \right|^2  
\; + \; \frac{\gamma}{4} \left| \Phi_{0,2} \Phi_{0,3} - e^{-2i \pi \beta_{23}} \Phi_{0,3} \Phi_{0,2} \right |^2    \nonumber \\
& + & \frac{\gamma}{4} \left | \Phi_{0,3} \Phi_{0,1} - e^{-2i \pi \beta_{31}} \Phi_{0,1} \Phi_{0,3} \right |^2 
\; + \;  \frac{\tilde{\gamma}}{4} \left \{ [\Phi_{0,1} , \Phi^{\dagger}_{0,1}] + [\Phi_{0,2} , \Phi^{\dagger}_{0,2}] + [\Phi_{0,3} , \Phi^{\dagger}_{0,3}] \right\}^2  \nonumber \\
& = &  \frac{\gamma}{4} \left | \Phi_{0,1} \Phi_{0,2} - e^{-2i \pi \beta_{12}} \Phi_{0,2} \Phi_{0,1} \right |^2  
\; + \; \frac{\gamma}{4} \left | \Phi_{0,2} \Phi_{0,3} - e^{-2i \pi \beta_{23}} \Phi_{0,3} \Phi_{0,2} \right |^2  \nonumber \\
& + & \frac{\gamma}{4} \left | \Phi_{0,3} \Phi_{0,1} - e^{-2i \pi \beta_{31}} \Phi_{0,1} \Phi_{0,3} \right |^2 ,
\label{mdnfoursusy}
\end{eqnarray}
where to obtain the last line in (\ref{mdnfoursusy}) we have used the fact that for untwisted fields  
\begin{eqnarray}
 [\Phi_{0,1} (x) , \Phi^{\dagger}_{0,1} (x)] & = & [\Phi_{0,2} (x) , \Phi^{\dagger}_{0,2} (x)] \quad = \quad [\Phi_{0,3} (x) , \Phi^{\dagger}_{0,3} (x)] \quad = \quad 0.
\label{sterm}
\end{eqnarray}
We can further simplify (\ref{mdnfoursusy}) and get 
\begin{eqnarray}
 \mathcal{H}_{0, \rm{int}} & = &   \frac{\gamma}{2}(1 - \cos 2\sigma_{12})\,\Phi^{\dagger}_{0,1} \Phi^{\dagger}_{0,2} \Phi_{0,1} \Phi_{0,2}  
\; + \;  \frac{\gamma}{2}(1 - \cos 2\sigma_{23})\,\Phi^{\dagger}_{0,2} \Phi^{\dagger}_{0,3}  \Phi_{0,2} \Phi_{0,3}  \nonumber \\
& + &  \frac{\gamma}{2} (1 - \cos 2\sigma_{31}) \, \Phi^{\dagger}_{0,3} \Phi^{\dagger}_{0,1} \Phi_{0,3} \Phi_{0,1} \nonumber \\
& = &  -\tilde{\gamma}' \sin^2 \sigma_{12} \: \Phi^{\dagger}_{0,1} \Phi^{\dagger}_{0,2} \Phi_{0,1} \Phi_{0,2}  
\; - \; \tilde{\gamma}' \sin^2 \sigma_{23} \: \Phi^{\dagger}_{0,2} \Phi^{\dagger}_{0,3} \Phi_{0,2} \Phi_{0,3}  
\; - \; \tilde{\gamma}' \sin^2 \sigma_{31} \: \Phi^{\dagger}_{0,3} \Phi^{\dagger}_{0,1} \Phi_{0,3} \Phi_{0,1} \nonumber \\
& = &  \frac{\gamma_{12}}{2}\: \Phi^{\dagger}_{0,1} \Phi^{\dagger}_{0,2} \Phi_{0,1} \Phi_{0,2}  \; + \; \frac{\gamma_{23}}{2} \: \Phi^{\dagger}_{0,2} \Phi^{\dagger}_{0,3} \Phi_{0,2} \Phi_{0,3} 
 \; + \; \frac{\gamma_{31}}{2} \: \Phi^{\dagger}_{0,3} \Phi^{\dagger}_{0,1} \Phi_{0,3} \Phi_{0,1} \nonumber \\
& = &  \mathcal{H}_{\theta, \rm{int}},
\label{mdn4susy}
\end{eqnarray}
where we have identified $2\sigma_{rs} =  - 2\pi\beta_{rs} = \lambda^{(r)}\wedge \lambda^{(s)}$ and $\tilde{\gamma}' = -\gamma$. Since $\sigma_{rs}$ 
and $\lambda^{(r)}\wedge \lambda^{(s)}$ are arbitrary parameters (due to arbitrariness of the components of $\theta$ matrix and of $\beta_{rs}$) so the above demands 
can be always satisfied. Hence, we infer that twisted scalar interaction Hamiltonian density (\ref{tgaugen4susy}) is equivalent to untwisted marginally deformed 
scalar interaction Hamiltonian density of (\ref{mdnfoursusy}). The marginal deformations of the $\mathcal{N} = 4$ SUSY theory arises in a natural way 
in the context of twisted field theories and it provides a general framework to discuss such theories.
Interested readers can also look at \cite{Beisert:2005if} for related work.


\section{Generic Twists}


So far we have restricted our discussion to only a very particular type of deformed statistics, which we called ``antisymmetric twisted statistics''. 
Such a twist is characterized by the the commutation relations (\ref{atwistedcomrel}) and (\ref{antitwistfield}). We called it an antisymmetric twist because the 
$\theta$ matrix characterizing it was a antisymmetric matrix. Now we want to discuss twists which are more general in nature. 

Let us consider a more general dressing transformation, which is  
\begin{eqnarray}
 a^R_r & = &   c_r \, e^{-\frac{i}{2} \lambda^{(r)}_l \tilde{\theta}^{(r)}_{lm} Q_m} ,  \nonumber \\
(a_r^R)^\dagger & = &   e^{\frac{i}{2} \lambda^{(r)}_l (\tilde{\theta}^{(r)})^\ast_{lm} Q_m} \, c^\dagger_r , \nonumber \\
 b^R_r  & = &  d_r \,  e^{\frac{i}{2} \lambda^{(r)}_l (\tilde{\theta}^{(r)})^\ast_{lm} Q_m} , \nonumber \\
(b_r^R)^\dagger  & = &  e^{-\frac{i}{2} \lambda^{(r)}_l \tilde{\theta}^{(r)}_{lm} Q_m} \, d^\dagger_r . 
\label{dresstranformmrmg}
\end{eqnarray}
Here $\tilde{\theta}^{(r)}$ is an arbitrary matrix and is not same for all particle species i.e.  $\tilde{\theta}^{(r)}_{lm} \neq \tilde{\theta}^{(s)}_{lm}$. 
Also, we have put a subscript $R$ to distinguish these transformations from the other possible transformations which are
%
%
\begin{eqnarray}
 a^L_r & = &    e^{-\frac{i}{2} \lambda^{(r)}_l \tilde{\theta}^{(r)}_{lm} Q_m}  \, c_r , \nonumber \\
(a_r^L)^\dagger & = &  c^\dagger_r \,  e^{\frac{i}{2} \lambda^{(r)}_l (\tilde{\theta}^{(r)})^\ast_{lm} Q_m},   \nonumber \\
 b^L_r  & = &    e^{\frac{i}{2} \lambda^{(r)}_l (\tilde{\theta}^{(r)})^\ast_{lm} Q_m} \, d_r , \nonumber \\
(b_r^L)^\dagger  & = &  d^\dagger_r \, e^{-\frac{i}{2} \lambda^{(r)}_l \tilde{\theta}^{(r)}_{lm} Q_m}.  
\label{dresstranformmlmg}
\end{eqnarray}  
Unlike the antisymmetric twist case, these two transformations are not equivalent but are related to each other as 
%
 \begin{eqnarray}
 a^R_r & = &   c_r \, e^{-\frac{i}{2} \lambda^{(r)}_l \tilde{\theta}^{(r)}_{lm} Q_m}   
\, = \,  e^{-\frac{i}{2} \lambda^{(r)}_l \tilde{\theta}^{(r)}_{lm} \lambda^{(r)}_m} \,  e^{-\frac{i}{2} \lambda^{(r)}_l \tilde{\theta}^{(r)}_{lm} Q_m}  \, c_r  
\, = \,e^{-\frac{i}{2} \lambda^{(r)}_l \tilde{\theta}^{(r)}_{lm} \lambda^{(r)}_m} \,  a^L_r .
\label{mrelarnal}
\end{eqnarray}
Similar relations hold for all other operators. 

Now, we consider the operator $N^R_r = (a_r^R)^\dagger a^R_r $ 
\begin{eqnarray}
N^R_r & = & (a_r^R)^\dagger a^R_r \, = \, e^{\frac{i}{2} \lambda^{(r)}_l (\tilde{\theta}^{(r)})^\ast_{lm} Q_m}\, c^\dagger_r\, c_r 
\, e^{-\frac{i}{2} \lambda^{(r)}_l \tilde{\theta}^{(r)}_{lm} Q_m}
\, = \,  c^\dagger_r\, c_r \, e^{\frac{i}{2}  \lambda^{(r)}_l ( (\tilde{\theta}^{(r)})^\ast_{lm} \, - \, \tilde{\theta}^{(r)}_{lm} ) Q_m}.
\label{genrnumberopm}
\end{eqnarray}
We restrict to the case of $N^R_r = c^\dagger_r\, c_r \, = \, N_{0,r}$, so that the twisted free Hamiltonian is equivalent to the untwisted one. 
The above condition implies that $(\tilde{\theta}^{(r)})^\ast_{lm} \, = \, \tilde{\theta}^{(r)}_{lm} $ i.e. all elements of  $\tilde{\theta}^{(r)}$ are real. 
Similar conditions hold for left twists. 

Also, we introduce the compact notation 
\begin{eqnarray}
\lambda^{(r)}_l \theta^{(r)}_{lm}  & = & 2 \alpha^{(r)}_m.
\label{veeproductm}
\end{eqnarray}
So the dressing transformations of (\ref{dresstranformmrmg}) and (\ref{dresstranformmlmg}) in this notation become
\begin{eqnarray}
 a^R_r & = &   c_r \, e^{-i \alpha^{(r)}_m Q_m} \, \equiv \, c_r \, e^{-i \alpha^{(r)}  Q} ,   \nonumber \\
(a_r^R)^\dagger & = &  e^{i \alpha^{(r)}_m Q_m} \, c^\dagger_r \, \equiv \, e^{i \alpha^{(r)} Q} \, c^\dagger_r , \nonumber \\
 b^R_r  & = &  d_r \, e^{i \alpha^{(r)}_m Q_m}  \, \equiv \,  d_r \, e^{i \alpha^{(r)} Q} , \nonumber \\
(b_r^R)^\dagger  & = & e^{-i \alpha^{(r)}_m Q_m} \, d^\dagger_r  \, \equiv \, e^{-i \alpha^{(r)} Q} \, d^\dagger_r , \nonumber \\
 a^L_r & = &   e^{-i \alpha^{(r)}_m Q_m}  \, c_r  \, \equiv \, e^{-i \alpha^{(r)} Q}  \, c_r , \nonumber \\
(a_r^L)^\dagger & = &  c^\dagger_r \, e^{i \alpha^{(r)}_m Q_m}  \, \equiv \,  c^\dagger_r \, e^{i \alpha^{(r)} Q} ,  \nonumber \\
 b^L_r  & = &  e^{i \alpha^{(r)}_m Q_m} \, d_r  \, \equiv \,  e^{i \alpha^{(r)} Q} \, d_r ,  \nonumber \\
(b_r^L)^\dagger  & = &  d^\dagger_r \, e^{-i \alpha^{(r)}_m Q_m}  \, \equiv \,   d^\dagger_r \, e^{-i \alpha^{(r)} Q}. 
\label{dresstranformalpha}
\end{eqnarray}  
The creation/annihilation operators defined by (\ref{dresstranformalpha}) satisfy twisted statistics of the form
\begin{eqnarray}
 a^R_r \, a^R_s & = &   c_r \, e^{-i \alpha^{(r)} Q}  \,  c_s \, e^{-i \alpha^{(s)} Q}   
\, = \, e^{ i \alpha^{(r)}\lambda^{(s)}} \,c_r \,c_s \, e^{-i \alpha^{(s)} Q} \,  e^{-i \alpha^{(r)} Q}  \nonumber \\
& = & \eta \, e^{ i \alpha^{(r)}\lambda^{(s)}} \,c_s \,c_r \, e^{-i \alpha^{(s)} Q} \,  e^{-i \alpha^{(r)} Q} \nonumber \\
& = & \eta \, e^{ i \alpha^{(r)}\lambda^{(s)}} \, e^{ - i \alpha^{(s)}\lambda^{(r)}}  \,c_s \, e^{-i \alpha^{(s)} Q} \,c_r \, e^{-i \alpha^{(r)} Q} \nonumber \\
& = & \eta \, e^{\frac{i}{2} \lambda^{(r)}_l \theta^{(r)}_{lm} \lambda^{(s)}_m} \, e^{-\frac{i}{2} \lambda^{(s)}_l \theta^{(s)}_{lm} \lambda^{(r)}_m} \, a_s \, a_r \nonumber \\
& = & \eta \, e^{\frac{i}{2} \lambda^{(r)}_l \theta^{(r)}_{lm} \lambda^{(s)}_m} \, e^{-\frac{i}{2} \lambda^{(r)}_l \theta^{(s)}_{ml} \lambda^{(s)}_m} \, a_s \, a_r \nonumber \\
& = & \eta \, e^{\frac{i}{2} \lambda^{(r)}_l \left(\theta^{(r)}_{lm} - \theta^{(s)}_{ml} \right) \lambda^{(s)}_m} \, a_s \, a_r .
\label{mstwiststat}
\end{eqnarray}
Since, $\theta^{(r)}_{lm} \neq \theta^{(s)}_{lm}$, and the $\theta$s are arbitrary matrices so $2 \theta'_{lm} = \theta^{(r)}_{lm} - \theta^{(s)}_{ml}$ also remains an arbitrary matrix
and hence (\ref{mstwiststat}) gives more general twisted statistics. Also for twisted bosons $\eta = 1$ should be taken and for twisted fermions $\eta = -1$ is to be taken. 

Similarly we find that 
\begin{eqnarray}
(a^R_r)^{\dagger} \, (a^R_s)^{\dagger} & = & \eta \, e^{ i \left( \alpha^{(r)}\lambda^{(s)} -  \alpha^{(s)}\lambda^{(r)} \right)} \, (a^R_s)^{\dagger} \, (a^R_r)^{\dagger}  
\, = \, \eta \, e^{\frac{i}{2} \lambda^{(r)}_l \left(\theta^{(r)}_{lm} - \theta^{(s)}_{ml} \right) \lambda^{(s)}_m}  \, (a^R_s)^{\dagger} \, (a^R_r)^{\dagger}, \nonumber \\
b^R_r \, b^R_s & = & \eta \, e^{ i \left( \alpha^{(r)}\lambda^{(s)} -  \alpha^{(s)}\lambda^{(r)} \right)} \, b^R_s \, b^R_r  
\, = \, \eta \, e^{\frac{i}{2} \lambda^{(r)}_l \left(\theta^{(r)}_{lm} - \theta^{(s)}_{ml} \right) \lambda^{(s)}_m}  \, b^R_s \, b^R_r  , \nonumber \\
(b^R_r)^{\dagger} \, (b^R_s)^{\dagger} & = & \eta \, e^{ i \left( \alpha^{(r)}\lambda^{(s)} -  \alpha^{(s)}\lambda^{(r)} \right)} \, (b^R_s)^{\dagger} \, (b^R_r)^{\dagger}  
\, = \, \eta \, e^{\frac{i}{2} \lambda^{(r)}_l \left(\theta^{(r)}_{lm} - \theta^{(s)}_{ml} \right) \lambda^{(s)}_m}  \, (b^R_s)^{\dagger} \, (b^R_r)^{\dagger} , \nonumber \\
 a^R_r \,  (a^R_s)^{\dagger} & = & \eta \, e^{ - i \left( \alpha^{(r)}\lambda^{(s)} -  \alpha^{(s)}\lambda^{(r)} \right)}   \, (a^R_s)^{\dagger} \,  a^R_r  
\,  + \, (2\pi)^3 \, 2 E_p \, \delta_{rs} \, \delta^{3}(p_{1} \, - \, p_{2}) \nonumber \\
& = & \, \eta \, e^{-\frac{i}{2} \lambda^{(r)}_l \left(\theta^{(r)}_{lm} - \theta^{(s)}_{ml} \right) \lambda^{(s)}_m}  \, (a^R_s)^{\dagger} \,  a^R_r  
\,  + \, (2\pi)^3 \, 2 E_p \, \delta_{rs} \, \delta^{3}(p_{1} \, - \, p_{2}) ,\nonumber \\
b^R_r \, (b^R_s)^{\dagger} & = & \eta \, e^{ -i \left( \alpha^{(r)}\lambda^{(s)} -  \alpha^{(s)}\lambda^{(r)} \right)} \, (b^R_s)^{\dagger} \, b^R_r
+ (2\pi)^3 \, 2 E_p \, \delta_{rs} \, \delta^{3}(p_{1} \, - \, p_{2}) \nonumber \\
& = & \eta \, e^{-\frac{i}{2} \lambda^{(r)}_l \left(\theta^{(r)}_{lm} - \theta^{(s)}_{ml} \right) \lambda^{(s)}_m} \,  (b^R_s)^{\dagger} \, b^R_r
+ (2\pi)^3 \, 2 E_p \, \delta_{rs} \, \delta^{3}(p_{1} \, - \, p_{2}), \nonumber \\
a^R_r \, (b^R_s)^{\dagger} & = & \eta \, e^{ i \left( \alpha^{(r)}\lambda^{(s)} -  \alpha^{(s)}\lambda^{(r)} \right)} \, (b^R_s)^{\dagger} \,  a^R_r
\, = \, \eta \, e^{\frac{i}{2} \lambda^{(r)}_l \left(\theta^{(r)}_{lm} - \theta^{(s)}_{ml} \right) \lambda^{(s)}_m}  \, (b^R_s)^{\dagger} \,  a^R_r ,\nonumber \\
a^R_r \, b^R_s & = & \eta \, e^{ - i \left( \alpha^{(r)}\lambda^{(s)} -  \alpha^{(s)}\lambda^{(r)} \right)} \, b^R_s \, a^R_r
\, = \, \eta \, e^{- \frac{i}{2} \lambda^{(r)}_l \left(\theta^{(r)}_{lm} - \theta^{(s)}_{ml} \right) \lambda^{(s)}_m}  \, b^R_s \, a^R_r.
\label{mstwistedcomrel}
\end{eqnarray}
From (\ref{mstwistedcomrel}) it is clear that the antisymmetric twist discussed in the previous section is just a special case of this generic twist. If we take 
$\theta^{(r_1)} = \theta^{(r_2)} = \cdots = \theta^{(r_N)} = \theta$ and $\theta_{lm} = - \theta_{ml}$ in (\ref{mstwistedcomrel}), we will recover back the antisymmetric twisted
statistics of the previous section. Moreover, unlike the case of antisymmetric twist, where due to antisymmetry of the $\theta$ matrix, it was not possible to get twisted 
statistics for an internal symmetry group of rank less than 2, in this case, we can have twisted statistics for SU(2) as well as U(1) group.

Using the twisted creation/annihilation operators, the left and right twisted quantum fields $\phi^{L,R}_{\theta,r}$ can be composed as \footnote{One can also compose fields with 
only left twisted or right twisted creation/annihilation operators but field theories with such quantum fields are tricky to write and one has to introduce quantities like 
``complex mass'' to write such theories. We will not discuss them in this work. }
\begin{eqnarray}
\phi^R_{\theta, r} (x) & = & \int \frac{d^3 p}{(2\pi)^3} \frac{1}{2E_{p}} \left[ a^R_r(p) e^{-i px} \, + \, (b^L_r)^{\dagger}(p) e^{i px} \right ] ,\nonumber \\
\phi^L_{\theta, r} (x) & = & \int \frac{d^3 p}{(2\pi)^3} \frac{1}{2E_{p}} \left[ a^L_r(p) e^{-i px} \, + \, (b^R_r)^\dagger (p) e^{i px} \right ], \nonumber \\
(\phi^R_{\theta, r})^\dagger (x) & = & \int \frac{d^3 p}{(2\pi)^3} \frac{1}{2E_{p}} \left[ b^L_r(p) e^{-i px} \, + \, (a^R_r)^{\dagger}(p) e^{i px} \right ], 
\nonumber 
\end{eqnarray}
\begin{eqnarray}
(\phi^L_{\theta, r})^\dagger (x) & = & \int \frac{d^3 p}{(2\pi)^3} \frac{1}{2E_{p}} \left[ b^R_r(p) e^{-i px} \, + \, (a^L_r)^\dagger (p) e^{i px} \right ]. 
\label{modeexpansionagen}
\end{eqnarray}

Using the dressing transformations of (\ref{dresstranformalpha}), it is easy to check that the above defined fields satisfy the dressing transformations 
\begin{eqnarray}
\phi^R_{\theta, r} (x) & = &  \phi_{0, r} (x) \, e^{-i \alpha^{(r)}  Q},  \nonumber \\
\phi^L_{\theta, r} (x) & = &  e^{-i \alpha^{(r)}  Q} \, \phi_{0, r} (x), \nonumber \\
(\phi^R_{\theta, r})^\dagger (x) & = & e^{i \alpha^{(r)}  Q} \, \phi^\dagger_{0, r} (x), \nonumber \\
(\phi^L_{\theta, r})^\dagger (x) & = &  \phi^\dagger_{0, r} (x) \, e^{i \alpha^{(r)}  Q}. 
\label{dressfieldsg}
\end{eqnarray}
Also, we have 
\begin{eqnarray}
 \left[ Q_m , \phi^{L,R}_{\theta, r} (x) \right ] & = & -\lambda^{(r)}_m \phi^{L,R}_{\theta, r} (x),  \nonumber \\
 \left[ Q_m , (\phi^{L,R}_{\theta, r})^{\dagger} (x) \right ] & = &  \lambda^{(r)}_m (\phi^{L,R}_{\theta, r})^{\dagger} (x).
\label{chargegen}
\end{eqnarray}
As before, the Fock space states can be constructed using these twisted operators. We assume (with similar justification as for the case of antisymmetric twists) that the vacuum 
of the twisted theory is same as that for untwisted theory. The multi-particle states can be obtained by acting the twisted creation operators on the vacuum state. 
Because of the twisted statistics (\ref{mstwistedcomrel}), there is an ambiguity in defining the action of the twisted creation and annihilation operators on Fock space states.
Like the previous case, we choose to define $a^\dagger_r (p)$, $p$ being the momentum label and $a^\dagger_r$ standing for either of the left or right twisted creation operators, 
to be an operator which adds a particle to the right of the particle list i.e.
\begin{eqnarray}
a^\dagger_r (p) | p_1,r_1; \, p_2,r_2;\, \dots \, p_n,r_n \rangle_{\theta} & = & | p_1,r_1;\, p_2,r_2;\, \dots \, p_n,r_n;\, p,r \rangle_{\theta} .
\label{actiontcre}
\end{eqnarray}
Again, it should be remarked that the particular choice (\ref{actiontcre}) is just a convention and we could have chosen the other convention where $a^\dagger_r (p)$
adds a particle to the left of the particle list. The two choices are not independent but are related to each other by a phase. Furthermore, the other choice can at most result in 
an overall phase in the $S$-matrix elements which will not have any observable effect.
With this convention, the single-particle Fock space states for the twisted theory are given by 
\begin{eqnarray}
\overline{|p, r \rangle}_{\theta} & = &   b^\dagger_r (p) | 0 \rangle , \nonumber \\
|p, r \rangle_{\theta} & = &  a^\dagger_r (p) | 0 \rangle.
\label{gsat}
\end{eqnarray}
The multi-particle states are given by
\begin{eqnarray}
\overline{| p_1, r_1; \, p_2,r_2; \, \dots \, p_n,r_n \rangle}_{\theta} & = &  b^\dagger_{r_n} (p_n) \dots b^\dagger_{r_2} (p_2) b^\dagger_{r_1} (p_1) | 0 \rangle ,\nonumber \\
|p_1, r_1; \, p_2,r_2; \, \dots \, p_n,r_n \rangle_{\theta} & = &  a^\dagger_{r_n} (p_n) \dots a^\dagger_{r_2} (p_2) a^\dagger_{r_1} (p_1) | 0 \rangle .
\label{gmat}
\end{eqnarray}
Owing to the twisted commutation relations of (\ref{mstwistedcomrel}), the state vectors also satisfy a similar twisted relation e.g. for two-particle states we have 
\begin{eqnarray}
\overline{| p_2, r_2; \, p_1,r_1 \rangle}_{\theta} & = & e^{ i \left( \alpha^{(r)}\lambda^{(s)} - \alpha^{(s)}\lambda^{(r)} \right)} 
\,\overline{| p_1, r_1; \, p_2,r_2 \rangle}_{\theta}, \nonumber \\
 |p_2, r_2; \, p_1,r_1 \rangle_{\theta} & = &  e^{ i \left( \alpha^{(r)}\lambda^{(s)} -  \alpha^{(s)}\lambda^{(r)} \right)}  \, |p_1, r_1; \, p_2,r_2 \rangle_{\theta} . 
\label{g2statecom}
\end{eqnarray}
The $SU(N)$ transformations of the twisted fields can be discussed in a way similar to the previous section. For example, the fields $\phi^L_{\theta, r}$ 
transform under $SU(N)$ as
\begin{eqnarray}
U(\sigma) \phi^L_{\theta, r} (x) U^\dagger(\sigma) & = &  \phi^{L'}_{\theta, r} (x) \, = \,\, U(\sigma) e^{-i \alpha^{(r)}  Q} \, U^\dagger(\sigma) \,
U(\sigma) \phi_{0, r} (x) U^\dagger(\sigma) \nonumber \\ 
& = & U(\sigma) e^{-i \alpha^{(r)}  Q} U^\dagger(\sigma) \, \left( e^{-i \sigma_a T_a} \right)_{rs} \,\phi_s (x) \, 
\, = \, \zeta_{(r)}(\sigma) \, \left( e^{-i \sigma_a T_a} \right)_{rs} \,\phi_s (x) , \nonumber \\
U(\sigma) (\phi^L_{\theta, r})^\dagger (x) U^\dagger(\sigma) & = & (\phi^{L'}_{\theta, r})^\dagger  (x) \, = \, U(\sigma) \phi^\dagger_{0, r} (x) U^\dagger(\sigma) 
\, U(\sigma) e^{i \alpha^{(r)}  Q} U^\dagger(\sigma) \nonumber \\ 
& = & \left( e^{i \sigma_a T_a} \right)_{sr} \,\phi^\dagger_s (x) \, U(\sigma)  e^{i \alpha^{(r)}  Q} U^\dagger(\sigma) 
\, = \,\left( e^{i \sigma_a T_a} \right)_{sr} \,\phi^\dagger_s (x) \, \zeta^\dagger_{(r)}(\sigma), 
\label{suntranformgen}
\end{eqnarray}
where $\zeta_{(r)}(\sigma) = U(\sigma)  e^{-i \alpha^{(r)}  Q} U^\dagger(\sigma)$ is an unitary operator satisfying 
$\zeta_{(r)}(\sigma)\zeta^\dagger_{(r)}(\sigma) =  \zeta^\dagger_{(r)}(\sigma) \zeta_{(r)}(\sigma) = \textbf{I}$. Similar relations hold for $\phi^R_{\theta, r}$ fields also.

The transformation properties of the state vectors can be similarly discussed. For example, assuming that vacuum remains invariant under the transformations 
i.e. $ U(\sigma) | 0 \rangle = | 0 \rangle$, the single-particle states transform as
\begin{eqnarray}
 U(\sigma) | r \rangle_\theta & = &  U(\sigma) a^\dagger_r | 0 \rangle \, = \, U(\sigma) a^\dagger_r U^\dagger(\sigma) U(\sigma) | 0 \rangle 
 \, = \, U(\sigma) c^\dagger_r \, e^{i \alpha^{(r)}  Q} \, U^\dagger(\sigma) U(\sigma) | 0 \rangle \nonumber \\ 
& = &  U(\sigma) c^\dagger_r U^\dagger(\sigma)| 0 \rangle \, = \, \left( e^{i \sigma_a T_a} \right)_{sr} \,c^\dagger_s | 0 \rangle 
\, = \, \left( e^{i \sigma_a T_a} \right)_{sr} \,a^\dagger_s | 0 \rangle \, = \, \left( e^{i \sigma_a T_a} \right)_{sr}  | s \rangle_\theta, \nonumber \\
 U(\sigma) \overline{| r \rangle}_\theta & = &  U(\sigma) b^\dagger_r | 0 \rangle \, = \, U(\sigma) b^\dagger_r U^\dagger(\sigma) U(\sigma) | 0 \rangle 
 \, = \, U(\sigma) d^\dagger_r \, e^{-i \alpha^{(r)}  Q} \, U^\dagger(\sigma) U(\sigma) | 0 \rangle \nonumber \\
& = & U(\sigma) d^\dagger_r U^\dagger(\sigma)| 0 \rangle \, = \, \left( e^{-i \sigma_a T^\ast_a} \right)_{sr} \,d^\dagger_s | 0 \rangle
\, = \, \left( e^{-i \sigma_a T^\ast_a} \right)_{sr} \,b^\dagger_s | 0 \rangle  \, = \, \left( e^{-i \sigma_a T^\ast_a} \right)_{sr}  \overline{| s \rangle}_\theta.
\label{1tstesuntrans}
\end{eqnarray}
Again, the multi-particle states follow twisted transformation rules, e.g. the two-particle states transform as
\begin{eqnarray}
 U(\sigma) | r , s \rangle_\theta & = & e^{i \alpha^{(s)} \lambda^{(r)}} \, e^{-i \alpha^{(t)} \lambda^{(u)}} \, \left( e^{i \sigma_a T_a} \right)_{ts}  
\left( e^{i \sigma_a T_a} \right)_{ur} \, | u ,t \rangle_\theta ,\nonumber \\
U(\sigma) \overline{| r, s \rangle}_\theta & = & e^{i \alpha^{(s)} \lambda^{(r)}} \, e^{-i \alpha^{(t)} \lambda^{(u)}}  \, \left( e^{-i \sigma_a T^\ast_a} \right)_{ts}\, 
\left( e^{-i \sigma_a T^\ast_a} \right)_{ur}\, \overline{| u, t \rangle}_\theta.
\label{2atstatvecsuntrans}
\end{eqnarray}
Since the left and right twisted fields have analogous properties, so henceforth we will consider only left twisted fields and will drop the superscript `` L '' from it. 
All the computations and conclusions applicable to left twisted fields can be equally applied to right twisted fields.

Now we have to define the analogue of star-product of previous section. We define the ``generic star-product'' $\ast$ as 
\begin{eqnarray}
 \phi^\#_{\theta, r} (x) \ast \phi^\#_{\theta, s}(y) & = &  \phi^\#_{\theta, r} (x) \, e^{ \frac{i}{2} \left( \left( \pm \alpha^{(s)} \right) \overleftarrow{Q} 
- \left(\pm \alpha^{(r)} \right) \overrightarrow{Q} \right)} \, \phi^\#_{\theta, s} (y)\nonumber \\
& = &  \phi^\#_{\theta, r} (x) \phi^\#_{\theta, s} (y) \, + \, \frac{i}{2} \left\{ \left(\pm \alpha^{(s)}_l \right) \left[ Q_l, \phi^\#_{\theta, r} (x)\right] 
- \left(\pm \alpha^{(r)}_m \right) \left[ Q_m, \phi^\#_{\theta, s} (y) \right] \right \} \nonumber \\
& + & \frac{1}{2!} \left(\frac{i}{2}\right)^2 \, \left\{ \left(\pm \alpha^{(s)}_l \right) \left(\pm \alpha^{(s)}_n \right) \left[ Q_l, \left[ Q_n,\phi^\#_{\theta, r} (x)\right]\right] 
-  \left(\pm \alpha^{(r)}_m \right) \left(\pm \alpha^{(r)}_p \right) \left[ Q_m, \left[ Q_p,\phi^\#_{\theta, s} (y) \right] \right] \right \} \nonumber \\
& + & \cdots,
\label{starproduct}
\end{eqnarray}
where $ + \alpha^{(r)}$ is to be taken if the field $\phi^\#_{\theta, r}$ stands for  $\phi^\dagger_{\theta, r}$ and $ - \alpha^{(r)}$ 
if  $\phi^\#_{\theta, r}$ stands for  $\phi_{\theta, r}$. 

Due to the relation (\ref{chargegen}), we have 
\begin{eqnarray}
 \phi^\#_{\theta, r} (x) \ast \phi^\#_{\theta, s}(y) & = &   e^{ \frac{i}{2} \left( \left( \pm \alpha^{(s)}\right) \left( \pm \lambda^{(r)}\right) 
- \left( \pm \alpha^{(r)}\right) \left( \pm \lambda^{(s)}\right) \right)}  \,\phi^\#_{\theta, r} (x) \, \cdot \, \phi^\#_{\theta, s}(y),
\label{astproductrel}
\end{eqnarray}
where $\alpha^{(r)}$, $\lambda^{(r)}$ should be taken if $\phi^\#_{\theta, r} = \phi^\dagger_{\theta, r}$ and $ - \alpha^{(r)}$, $-\lambda^{(r)}$ should be taken if 
$\phi^\#_{\theta, r} = \phi_{\theta, r}$. So again the $\ast$-product is nothing but multiplying a certain phase factor to ordinary product. Nonetheless like in the 
previous case, using $\ast$-product will simplify things and the product should be regarded as a compact notation used for convenience. 

Having defined the product rule on single fields we have to now define how it acts on a composition of fields like $\left(\phi^\#_{\theta, r} \phi^\#_{\theta, s}\right)$.
Demanding that our product remains associative, the action of the $\ast$-product on composition of fields is defined as
\begin{eqnarray}
 \phi^\#_{\theta, r} \ast \left( \phi^\#_{\theta, s} \phi^\#_{\theta, t} \right) & = &  \phi^\#_{\theta, r} \, 
e^{ \frac{i}{2} \left( \left( \pm \alpha^{(s)} \pm \alpha^{(t)} \right) \overleftarrow{Q} - \left(\pm \alpha^{(r)} \right) \overrightarrow{Q} \right)} \,
\left( \phi^\#_{\theta, s} \phi^\#_{\theta, t} \right) \nonumber \\
& = &  \phi^\#_{\theta, r} \phi^\#_{\theta, s} \phi^\#_{\theta, t} \, + \, \frac{i}{2} \left\{ \left(\pm \alpha^{(s)}_l \pm \alpha^{(t)}_l \right) 
\left[ Q_l, \phi^\#_{\theta, r} \right] - \left(\pm \alpha^{(r)}_m \right) \left[ Q_m, \left( \phi^\#_{\theta, s} \phi^\#_{\theta, t} \right) \right] \right \} \nonumber \\
& + & \frac{1}{2!} \left(\frac{i}{2}\right)^2 \, \left\{ \left(\pm \alpha^{(s)}_l \pm \alpha^{(t)}_l \right) \left(\pm \alpha^{(s)}_n \pm \alpha^{(t)}_n \right)
 \left[ Q_l, \left[ Q_n,\phi^\#_{\theta, r} \right]\right] \right. \nonumber \\
& - & \left. \left(\pm \alpha^{(r)}_m \right) \left(\pm \alpha^{(r)}_p \right) \left[ Q_m, \left[ Q_p,\left( \phi^\#_{\theta, s} \phi^\#_{\theta, t} \right) \right] \right] \right \} 
\, + \, \cdots
\label{3starproduct}
\end{eqnarray}
Similar action of $\ast$-product applies for composition of multiple fields
\begin{eqnarray}
\left( \phi^\#_{\theta, r_1} \cdots \phi^\#_{\theta, r_N} \right) \ast \left( \phi^\#_{\theta, s_1} \cdots \phi^\#_{\theta, s_M} \right) 
& = & \left( \phi^\#_{\theta, r_1} \cdots \phi^\#_{\theta, r_N} \right) \, 
e^{\frac{i}{2} \left(\left(\pm \alpha^{(s_1)}\cdots \pm \alpha^{(s_M)} \right) \overleftarrow{Q} - \left(\pm \alpha^{(r_1)}\cdots \pm \alpha^{(r_N)} \right)\overrightarrow{Q} \right)} 
\nonumber \\
& & \left( \phi^\#_{\theta, s_1} \cdots \phi^\#_{\theta, s_M} \right). 
\label{mstarproduct}
\end{eqnarray}
The above defined $\ast$-product can be viewed as the internal space analogue of another widely studied spacetime product called ``Dipole Product'' \cite{dipole-keshav}. 

The star-product introduced here has the property that
\begin{eqnarray}
 \phi^\#_{\theta,r} \, \ast  \, \phi^\#_{\theta, s} & = &  \phi^\#_{\theta, s} \, \ast  \, \phi^\#_{\theta,r}, \nonumber \\
 \phi^\#_{\theta,r} \, \ast  \, \left( \phi^\#_{\theta, s} \, \ast  \, \phi^\#_{\theta, t} \right) 
& = & \left( \phi^\#_{\theta,r} \, \ast  \, \phi^\#_{\theta, s} \right)\, \ast  \, \phi^\#_{\theta, t}, 
\label{astidanti1}
\end{eqnarray}
and by construction we have
\begin{eqnarray}
 \phi^\#_{\theta,r} \, \ast  \, \phi^\#_{\theta, r} & = &  \phi^\#_{\theta, r} \cdot \phi^\#_{\theta,r} .
\label{astidanti2}
\end{eqnarray}
Also, because of the dressing transformations (\ref{dressfieldsg}) we have
\begin{eqnarray}
 \phi^\#_{\theta,r_1} \, \ast  \, \phi^\#_{\theta, r_2} \, \ast  \, \cdots \, \ast  \, \phi^\#_{\theta, r_n} & = & \kappa \, \phi^\#_{0,r_1} \, \phi^\#_{0, r_2} \, \cdots \, 
\phi^\#_{0, r_n} \, e^{i \left( \pm \alpha^{(r_1)} \, \pm \,  \alpha^{(r_2)} \, \pm \, \cdots \, \pm \,  \alpha^{(r_n)} \right) Q },
\label{astidanti3}
\end{eqnarray}
where $\kappa $ is a phase factor whose explicit form depends on whether  $ \phi^\#_{\theta,r}$ stands for $\phi^\dagger_{\theta,r}$ or $ \phi_{\theta,r}$ field.
Also, $ + \alpha^{(r)}$ is to be taken if $ \phi^\#_{\theta,r} = \phi^\dagger_{\theta,r}$ and $ -\alpha^{(r)}$ if $ \phi^\#_{\theta,r} = \phi_{\theta,r}$. 

Having defined the $\ast$-product, the field theories can be conveniently written by following the rule that, for any given untwisted Hamiltonian its twisted counterpart 
should be written by replacing the untwisted fields $\phi_{0,r}$ by the twisted fields $\phi_{\theta,r}$ and the ordinary product between fields by the $\ast$-product. 

Using the above rule, the free theory Hamiltonian density $\mathcal{H}_{\theta, F}$ in terms of twisted fields can be written as
 \begin{eqnarray}
  \mathcal{H}_{\theta, F} & = &  \Pi^{\dagger}_{\theta,r} \ast  \Pi_{\theta,r}  \, + \, (\partial_i \phi^{\dagger}_{\theta,r}) \ast (\partial^i \phi_{\theta,r})
 \, + \, m^2  \, \phi^{\dagger}_{\theta,r} \ast  \phi_{\theta,r} \nonumber \\
& = &  \Pi^{\dagger}_{\theta,r} \Pi_{\theta,r}  \, + \, (\partial_i \phi^{\dagger}_{\theta,r}) (\partial^i \phi_{\theta,r})  \, + \, m^2  \, \phi^{\dagger}_{\theta,r} \phi_{\theta,r},
\label{freetham}
 \end{eqnarray}
where $\Pi_{\theta,r}$ is the canonical conjugate of $\phi_{\theta,r}$; $r = 1, 2, \cdots, N$ and to obtain the last line in (\ref{freetham}) we have used (\ref{astidanti2}). 
The Hamiltonian density in (\ref{freetham}) is invariant under the $SU(N)$ global transformations which can be explicitly checked by using (\ref{suntranformgen}). 

The renormalizable $SU(N)$ invariant interaction Hamiltonian density is given by
\begin{eqnarray}
\mathcal{H}_{\theta,\rm{Int}} & = &  \frac{\gamma}{4} \, \phi^{\dagger}_{\theta,r}  \ast   \phi^{\dagger}_{\theta,s}  \ast  \phi_{\theta,r}  \ast  \phi_{\theta,s} \nonumber \\
& = & \frac{\gamma}{4}\, e^{ i \left( \alpha^{(s)}\lambda^{(r)} -  \alpha^{(r)}\lambda^{(s)} \right)} \, \phi^{\dagger}_{\theta,r} \phi^{\dagger}_{\theta,s} 
\phi_{\theta,r} \phi_{\theta,s},
\label{intgenham}
\end{eqnarray}
where $r,s = 1, 2, \cdots, N$. Using (\ref{suntranformgen}) one can check that (\ref{intgenham}) is indeed invariant under $SU(N)$ group. 
The presence of phases of the type $e^{ i \left( \alpha^{(s)}\lambda^{(r)} -  \alpha^{(r)}\lambda^{(s)} \right)} $ in (\ref{intgenham}) means that unlike the untwisted case,
the demand of $SU(N)$ invariance forces the various terms to have different couplings related with each other in a specific way. 

Again like the previous section, the discussion in this section can be generalized in a straightforward manner to include spinor fields and twisted fermions. For that we have to take
anticommuting creation/annihilation operators and the twisted fermions will again satisfy (\ref{mstwistedcomrel}) but with $\eta = -1$. Also, the above discussion can be easily 
generalized to higher dimensional representations of $SU(N)$ group as well as to other symmetry groups like $SO(N)$.


\subsection{$S$-matrix Elements}


Like the previous case of antisymmetric twists, here also by construction we have ensured that the twisted $SU(N)$ invariant free Hamiltonian density $\mathcal{H}_{\theta, F}$ 
remains same as the untwisted one i.e.
 \begin{eqnarray}
  \mathcal{H}_{\theta, F} & = &  \Pi^{\dagger}_{\theta,r} \ast \Pi_{\theta,r}  \, + \, (\partial_i \phi^{\dagger}_{\theta,r}) \ast(\partial^i \phi_{\theta,r})
 \, + \, m^2  \, \phi^{\dagger}_{\theta,r} \ast \phi_{\theta,r} \nonumber \\
& = &  \Pi^{\dagger}_{0,r} \Pi_{0,r}  \, + \, (\partial_i \phi^{\dagger}_{0,r}) (\partial^i \phi_{0,r}) \, + \, m^2  \, \phi^{\dagger}_{0,r} \phi_{0,r}
\, = \,  \mathcal{H}_{0,F} ,
\label{freegenhamrel}
 \end{eqnarray}
where to obtain the last line we have used the dressing transformation (\ref{dressfieldsg}) and the relation (\ref{astidanti3}). Similarly, the twisted $SU(N)$ invariant interacting
Hamiltonian density $\mathcal{H}_{\theta,\rm{ I}}$  can be shown to be same as the untwisted one
 \begin{eqnarray}
  \mathcal{H}_{\theta, \rm{I}} & = &  \Pi^{\dagger}_{\theta,r} \ast \Pi_{\theta,r}  \, + \, (\partial_i \phi^{\dagger}_{\theta,r}) \ast(\partial^i \phi_{\theta,r})
 \, + \, m^2  \, \phi^{\dagger}_{\theta,r} \ast \phi_{\theta,r} 
\, + \,  \frac{\gamma}{4} \, \phi^{\dagger}_{\theta,r}  \ast  \phi^{\dagger}_{\theta,s}  \ast \phi_{\theta,r}  \ast \phi_{\theta,s} \nonumber \\
& = &  \Pi^{\dagger}_{0,r} \Pi_{0,r}  \, + \, (\partial_i \phi^{\dagger}_{0,r}) (\partial^i \phi_{0,r}) \, + \, m^2  \, \phi^{\dagger}_{0,r} \phi_{0,r}
\, + \,  \frac{\gamma}{4} \, \phi^{\dagger}_{0,r}  \phi^{\dagger}_{0,s}  \phi_{0,r}  \phi_{0,s} 
 \, = \, \mathcal{H}_{0, \rm{I}} .
\label{intgenhamrel}
 \end{eqnarray}
Since the twisted in/out states also contain information about twisted statistics so we should again look at the $S$-matrix elements. For a typical $S$-matrix element, like for the 
scattering process of $\phi_{\theta, r} \phi_{\theta, s} \rightarrow \phi_{\theta, r} \phi_{\theta, s}$ we have
 \begin{eqnarray}
  S\left[ \phi_{\theta, r} \phi_{\theta, s} \rightarrow \phi_{\theta, r} \phi_{\theta, s}  \right] 
& = & \leftidx{_{out,\theta}}{\left \langle r s | r s  \right \rangle }{_{ \theta,in}} 
\, = \, \leftidx{_{\theta}}{\left \langle r s | S_\theta | r s \right \rangle }{_{ \theta}},
\label{smatgen}
 \end{eqnarray}
where $S_\theta = \mathcal{T}  \exp \left[-i\int^{\infty}_{-\infty} d^{4}z \mathcal{H}_{\theta, \rm{ Int}} (z) \right] $ is the $S$-operator and we have denoted the two-particle in and 
out states by $\left| r s \right \rangle_\theta = a^\dagger_s a^\dagger_r | 0 \rangle$. Because of (\ref{intgenhamrel}) we have 
\begin{eqnarray}
 S_\theta & = & \mathcal{T}  \exp \left[-i\int^{\infty}_{-\infty} d^{4}z \mathcal{H}_{\theta, \rm{Int}} (z) \right]  
\, = \, \mathcal{T}  \exp \left[-i\int^{\infty}_{-\infty} d^{4}z \mathcal{H}_{0, \rm{Int}} (z) \right]  \, = \, S_0.
\label{sopgen}
\end{eqnarray}
Also we have 
\begin{eqnarray}
 \left| r s \right \rangle_\theta & = & a^\dagger_s a^\dagger_r  \left| 0 \right\rangle 
\, = \, e^{i \alpha^{(s)} \lambda^{(r)} } \,  c^\dagger_s \, c^\dagger_r \,\left| 0 \right\rangle  
\, = \, e^{i \alpha^{(s)} \lambda^{(r)} } \, \left| r s \right \rangle_0.
\label{stategen}
\end{eqnarray}
Using (\ref{stategen}) and (\ref{sopgen}) we have
 \begin{eqnarray}
  S\left[ \phi_{\theta, r} \phi_{\theta, s} \rightarrow \phi_{\theta, r} \phi_{\theta, s}  \right] & = & 
\leftidx{_{0}}{\left \langle r s \left| \, e^{-i \alpha^{(s)} \lambda^{(r)} } \, S_0 \, e^{i \alpha^{(s)} \lambda^{(r)} } \, \right| r s \right \rangle }{_{ 0}}
\nonumber \\
& = & \leftidx{_{0}}{\left \langle r s | \, S_0 \, | r s \right \rangle }{_{ 0}} \nonumber \\
& = &  S\left[ \phi_{0, r} \phi_{0, s} \rightarrow \phi_{0, r} \phi_{0, s}  \right]. 
\label{smatgenrel}
\end{eqnarray}
So like the antisymmetric twist case, the $S$-matrix elements of twisted $SU(N)$ invariant theory are same as that of the untwisted $SU(N)$ invariant theory. 
One can equally regard a $SU(N)$ invariant $S$-matrix element as due to untwisted fields with local interaction terms or due to nonlocal twisted fields.
Hence it is difficult to distinguish between the two theories by doing a scattering experiment. 


Again dropping either the demand of invariance of vacuum or invariance of the interaction term under $SU(N)$ transformations, will result into twisted theories 
being different from the untwisted
ones. For example if we do not multiply fields with $\ast$-product then the twisted Hamiltonian density with quartic interactions can be written as   
 \begin{eqnarray}
  \mathcal{H}_{\theta} & = &  \Pi^{\dagger}_{\theta,r} \Pi_{\theta,r}  \, + \, (\partial_i \phi^{\dagger}_{\theta,r}) (\partial^i \phi_{\theta,r})
 \, + \, m^2  \, \phi^{\dagger}_{\theta,r} \phi_{\theta,r} 
\, + \,  \frac{\gamma}{4} \, \left[ \phi^{\dagger}_{\theta,r}  \phi^{\dagger}_{\theta,s} \phi_{\theta,r}  \phi_{\theta,s} 
\, + \, \phi^{\dagger}_{\theta,s} \phi^{\dagger}_{\theta,r} \phi_{\theta,r}  \phi_{\theta,s} 
\, + \,  \phi^{\dagger}_{\theta,s} \phi^{\dagger}_{\theta,r}   \phi_{\theta,s} \phi_{\theta,r} \right . \nonumber \\  
& + & \left. \phi^{\dagger}_{\theta,r}  \phi^{\dagger}_{\theta,s}   \phi_{\theta,s} \phi_{\theta,r}  \, + \, \cdots  \right ] \, + \, \rm{h. c}.
\label{genhamwithoutast}
\end{eqnarray}
where $\cdots$ represents the other 24 possible terms which we can write. Some of these 24 terms are equivalent to other terms but 
unlike the untwisted case not all of them are equal to each other. Moreover, this Hamiltonian has no $SU(N)$ symmetry and it maps to the marginally deformed Hamiltonian of the 
untwisted theory. 

The interaction Hamiltonian given by
\begin{eqnarray}
  \mathcal{H}_{\theta,\rm{Int}} & = &  \frac{\gamma}{4} \, \phi^{\dagger}_{\theta,r}  \ast  \phi^{\dagger}_{\theta,r}  \ast \phi^{\dagger}_{\theta,s}  \ast \phi_{\theta,s} 
\, + \, \rm{h.c},
\label{hamsunviogen}
\end{eqnarray}
is not even equivalent to any local untwisted Hamiltonian, and maps to 
\begin{eqnarray}
 \mathcal{H}_{0,\rm{Int}} & = &  \frac{\gamma}{4} \, \phi^{\dagger}_{0,r} \phi^{\dagger}_{0,r} \phi^{\dagger}_{0,s} \phi_{0,s} \, e^{ i \lambda^{(r)} \wedge Q }
\, + \, \rm{h.c},
\label{hamsunviogenrel}
\end{eqnarray}
which is nonlocal because of the presence of nonlocal operators $Q_m$. 

In this section and the preceding one we constructed field theories involving nonlocal fields having twisted statistics. We only restricted to a small subset of all such possible 
twisted theories. One can consider various generalizations of such twisted theories and we plan to discuss more of them in later works.


\section{Causality of Twisted Field Theories}


In this section we briefly address the issue of causality of the twisted quantum field theories. As it turns out, the twisted fields and hence the twisted field theories constructed out 
of them
are in general non-causal but inspite of that one can construct Hamiltonian densities like the $SU(N)$ invariant ones, which are causal and also satisfy cluster decomposition principle.


\subsection{Commutative Case}


For sake of completeness, we start with reviewing the discussion of causality in the untwisted case. Again we limit our discussion only to scalar fields but similar arguments
(with appropriate modifications) also hold for spinor fields and anti commuting operators. In the untwisted case, for complex scalar fields $\phi_{0,r}(x)$; $r = 1,2, \cdots N$ 
we have 
\begin{eqnarray}
i \Delta^{0}_{rs}(x-y) & = & \left< 0 \left|\left[\phi_{0, r}(x), \phi^{\dagger}_{0, s}(y) \right] \right| 0 \right> \nonumber \\
& = & - \delta_{rs}\int \frac{d^3 p}{(2\pi)^3} \frac{\sin\,p(x-y)}{E_{p}}.
\label{comspacecommutator}
\end{eqnarray}
It can be checked that for space like separations i.e. $(x - y)^2 < 0$ we have 
 \begin{eqnarray}
  i \Delta^{0}_{rs}(x-y) & = & \left< 0 \left|\left[\phi_{0, r}(x), \phi^{\dagger}_{0, s}(y) \right] \right| 0 \right> \; = \; 0 \qquad \text{for} \; (x-y)^2 \, < \, 0.
\label{comcausalitycon}
 \end{eqnarray}
Also, we have 
\begin{eqnarray}
 \left < 0 |[\phi_{0, r}(x), \phi_{0, s}(y) ]| 0 \right > \; = \; \left< 0 \left| [\phi^{\dagger}_{0, r}(x), \phi^{\dagger}_{0, s}(y)] \right|0 \right > \; = \; 0.
\label{comothercom}
\end{eqnarray}
Using (\ref{comcausalitycon}) and (\ref{comothercom}) it can be easily shown that any local operator which is a functional of $\phi_{0, r}$, $\phi^{\dagger}_{0, s}$ and 
their derivatives will also follow a similar relation e.g consider a generic local bilinear operator 
%
\begin{eqnarray}
\Xi^0(x) = \phi^{\dagger}_{0, r}(x) \xi_{rs}(x)\phi_{0, s}(x), 
\label{combilinearop}
\end{eqnarray}
where $\xi_{rs}(x)$ can be either a c-number valued function or a differential operator. Now, we have
\begin{eqnarray}
[\Xi^0(x) \,, \, \Xi^0(y)] & = & [\phi^{\dagger}_{0, r}(x) \xi_{rs}(x)\phi_{0, s}(x) \,,\, \phi^{\dagger}_{0,u} (y) \xi_{uv}(y)\phi_{0, v} (y) ] \nonumber \\
& = & \xi_{rs}(x) \xi_{uv}(y)\: \left\{ \phi^{\dagger}_{0, r}(x) \: i \delta_{su}\Delta_{su}(x-y)\:\phi_{0, v} (y) 
\; +\;  \phi^{\dagger}_{0, u} (y)\: (-i) \delta_{rv}\Delta_{rv}(y-x)\:\phi_{0, s}(x) \right\}
\nonumber \\
& = & 0 \qquad \text{for} \; (x-y)^2 \, < \, 0.
\label{comopcom}
\end{eqnarray}
Similarly it can be shown that the self commutator of other local operators (at two different spacetime labels) which are functional of $\phi_{0, r}$, $\phi^{\dagger}_{0, r}$ and 
their derivatives will
always vanish for $(x-y)^2 \, < \, 0 $. In particular its straight forward to see that the self commutator of a local Hamiltonian density at two different spacetime labels always 
vanish for $(x-y)^2 \, < \, 0 $ i.e.
\begin{eqnarray}
 [\mathcal{H}(x), \mathcal{H} (y)] & = & 0 \qquad \text{for} \; (x-y)^2 \, < \, 0.
\label{comhamcon}
\end{eqnarray}
 Hence, in untwisted theory, (\ref{comcausalitycon}) is a sufficient although not necessary condition for the theory to be causal.


\subsection{Twisted Case}


In the twisted case, the relation analogous to (\ref{comcausalitycon}) does not hold. So for twisted case we have
\begin{eqnarray}
i \Delta^{\theta}_{rs}(x-y) & = & \left< 0 \left|\left[\phi_{\theta, r}(x), \phi^{\dagger}_{\theta, s}(y) \right] \right| 0 \right> \nonumber \\
& = &  \left< 0 \left|\left[ e^{-i \alpha^{(r)}  Q} \, \phi_{0, r}(x) \,,\, \phi^{\dagger}_{0, s}(y) e^{i \alpha^{(s)}  Q}\,  \right] \right| 0 \right> 
\nonumber \\
& = &  \left< 0 \left| \left \{ e^{-i \alpha^{(r)}  Q} \, \left[ \phi_{0, r}(x)\,,\,\phi^{\dagger}_{0, s}(y) \right]  \,  e^{i \alpha^{(s)} Q}
\; + \;  \left[ e^{-i \alpha^{(r)}  Q} \,,\, \phi^{\dagger}_{0, s}(y)\right] \, \phi_{0, r}(x)\, e^{i \alpha^{(s)} Q}
\right. \right. \right. \nonumber \\
& + & \left. \left.\left.  \phi^{\dagger}_{0, s}(y) \, e^{-i \alpha^{(r)}  Q} \, \left[\phi_{0, r}(x) \,,\,  e^{i \alpha^{(s)} Q} \right]
\; + \; \phi^{\dagger}_{0, s}(y) \,\left[ e^{-i \alpha^{(r)}  Q} \,,\, e^{i \alpha^{(s)} Q} \right] \, \phi_{0, r}(x) \right\}\right| 0 \right>
\nonumber \\
& = &  \left< 0 \left|\left[ \phi_{0, r}(x)\,,\,\phi^{\dagger}_{0, s}(y) \right]\right| 0 \right> 
\; + \; \left< 0 \left| e^{-i \alpha^{(r)} \lambda^{(s)}} \, \phi^{\dagger}_{0, s}(y) \, \phi_{0, r}(x)\, \right| 0 \right> \nonumber \\
& + & \left< 0 \left| e^{i \alpha^{(s)}  \lambda^{(r)}} \, \phi^{\dagger}_{0, s}(y) \, e^{-i \alpha^{(r)}  Q} \, \phi_{0, r}(x) \, \right| 0 \right> 
\nonumber \\
& = & i \delta_{rs} \Delta^{0}_{rs}(x - y) \; + \;  \left\{ e^{-i \alpha^{(r)} \lambda^{(s)}} \, + \,  e^{i (\alpha^{(s)} + \alpha^{(r)} )\lambda^{(r)}} \right\}
 \left< 0 \left|\phi^{\dagger}_{0, s}(y) \phi_{0, r}(x)\right| 0 \right> ,
\label{twistedcomrel}
\end{eqnarray}
where in last two steps we have used the fact that $Q |0> = |0>$. Let us denote $ \mathbf{A} = \left< 0 \left|\phi^{\dagger}_{0, s}(y) \phi_{0, r}(x)\right| 0 \right> $. 
It can be easily seen that although $\Delta^{0}_{rs}(x - y) $ vanishes for space-like separation but $\mathbf{A}$ does not vanish. For example, let us take the special 
case of $(x^0 - y^0) = 0$ and $(\vec{x} - \vec{y}) = \vec{z}$. This is a special case of space like separation i.e in this case $(x -y)^2 < 0$. Therefore, we have
\begin{eqnarray}
 \mathbf{A} & = & \left< 0 \left|\phi^{\dagger}_{0, s}(y) \phi_{0, r}(x)\right| 0 \right> \nonumber \\
& = & \int \frac{d^3 p}{(2\pi)^3} \frac{1}{2E_p} \int \frac{d^3 q}{(2\pi)^3} \frac{1}{2E_q}
\left< 0 \left|\left( c^{\dagger}_s(q)e^{iqy} \, + \, d_s(q)e^{-iqy} \right)\left( c_r(p)e^{-ipx} \, + \, d^{\dagger}_r(p)e^{ipx} \right)\right| 0 \right> \nonumber \\
& = & \int \frac{d^3 p}{(2\pi)^3} \frac{1}{2E_p} \int \frac{d^3 q}{(2\pi)^3} \frac{1}{2E_q} \left<0\left|d_s(q)d^{\dagger}_r(p)\right|0\right> e^{ipx} e^{-iqy}
\nonumber \\
& = & \int \frac{d^3 p}{(2\pi)^3} \frac{1}{2E_p} \int \frac{d^3 q}{(2\pi)^3} \frac{1}{2E_q} (2\pi)^3 2E_p \delta_{rs} \delta^3(p - q) e^{ipx} e^{-iqy} \nonumber \\
& = & \delta_{rs} \int \frac{d^3 p}{(2\pi)^3} \frac{1}{2E_p} e^{-i\vec{p}\vec{z}} \qquad \qquad \text{for} \; (x^0 - y^0) = 0 \; \text{and} \; (\vec{x} - \vec{y}) = \vec{z}.
\label{twistA}
\end{eqnarray}
Going to polar coordinates we have 
\begin{eqnarray}
  \mathbf{A} & = & 2\pi \delta_{rs} \int^{\infty}_{0} \frac{d p}{(2\pi)^3} \frac{|\vec{p}|}{\sqrt{|\vec{p}|^2 + m^2}} \frac{\sin |\vec{p}||\vec{z}|}{|\vec{z}|} \nonumber \\
& = & \frac{m\delta_{rs}}{4\pi^2 |\vec{z}|}K_1(m|\vec{z}|) ,
\label{twistedA}
\end{eqnarray}
where $K_1$ is the Hankel function. Clearly $\mathbf{A}$ does not vanish for all space-like separations. So using (\ref{twistedA}) in (\ref{twistedcomrel}) we have
\begin{eqnarray}
i \Delta^{\theta}_{rs}(x-y) & = &  i \delta_{rs} \Delta^{0}_{rs}(x - y) \; + \; \left( e^{-i \alpha^{(r)} \lambda^{(s)}} \, + \, e^{i (\alpha^{(s)} + \alpha^{(r)} )\lambda^{(r)}} \right)
\mathbf{A},
\label{twistedcausal}
\end{eqnarray}
which for a particular special case of space-like separations i.e. $(x-y)^2 \, = \, z^2 < 0 $  and $(x^0 - y^0) = 0$ is 
\begin{eqnarray}
 i \Delta^{\theta}_{rs}(x-y) & = &  \left( e^{-i \alpha^{(r)} \lambda^{(s)}} \, + \, e^{i (\alpha^{(s)} + \alpha^{(r)} )\lambda^{(r)}} \right) 
\frac{m\delta_{rs}}{4\pi^2 |\vec{z}|}K_1(m|\vec{z}|) \neq 0 .
\label{twistedcausality}
\end{eqnarray}
Hence, unlike the untwisted case, the twisted fields don't commute for all space-like separations. An immediate consequence of (\ref{twistedcausality}) is that it
can no longer be guaranteed that, the self-commutator at different spacetime labels, of all operators which are functional of the twisted fields (or their derivatives) will vanish
for space-like separations. In particular, following computations similar to (\ref{comopcom}), it can be shown that the self-commutator of generic twisted bilinear operators 
$\Xi^\theta(x) = \hat{\phi}^{\dagger}_{\theta, r}(x) \xi_{rs}(x)\hat{\phi}_{\theta, s}(x)$ does not vanish for all space-like separations i.e.
 \begin{eqnarray}
  [\Xi^\theta(x)\, , \, \Xi^\theta(y)] & \neq & 0 \qquad \text{for all} \; (x-y)^2 \; < \: 0 .
\label{twistedbilinear}
 \end{eqnarray}
A similar result will follow for any generic operator formed from these twisted fields. But as remarked earlier, the condition (\ref{comcausalitycon}) (or its twisted analogue) is
just a sufficient condition and by no means it is a necessary condition. Infact even in untwisted case, (\ref{comcausalitycon}) is not satisfied by fermionic fields. Therefore
inspite of (\ref{twistedcausality}) it is still possible to construct twisted Hamiltonian densities which satisfy causality constraints
\begin{eqnarray}
[\hat{\mathcal{H}}(x)\, , \, \hat{\mathcal{H}}(y)] & = & 0 \qquad \text{for} \; (x-y)^2 \; < \: 0 .
\label{twistedcausalham}
\end{eqnarray} 
It is easy to see that the twisted $SU(N)$ invariant Hamiltonian densities of (\ref{intantihamrel}) and (\ref{intgenhamrel}) satisfy the causality condition. 
But a generic Hamiltonian density constructed out of twisted fields will not necessarily satisfy (\ref{twistedcausalham}). For example, the nonlocal 
Hamiltonian densities of (\ref{hamsunvioanti}) and (\ref{hamsunviogen}) are noncausal.


\section{Conclusions}


In this paper we discussed the possibility of having twisted statistics by deformation of internal symmetries. We constructed two such deformed statistics and discussed field 
theories for such deformed fields. We showed that both type of twisted quantum fields discussed in this paper, satisfy commutation relations different from the usual bosonic/fermionic 
commutation relations. Such twisted fields by construction (and in view of CPT theorem) are nonlocal in nature. We showed that inspite of the basic ingredient fields being nonlocal,  
it is possible to construct interaction Hamiltonians which satisfy cluster decomposition principle and are Lorentz invariant. 

We first discussed a specific type of twist called ``antisymmetric twist''. This kind of twist is quite similar in spirit to the twisted noncommutative field theories.
The formalism developed for antisymmetric twists was analogous (with appropriate generalizations and modifications) to the formalism of twisted noncommutative theories. 
We then constructed interaction terms using such twisted fields and discussed the scattering problem for such theories. We found that the twisted $SU(N)$ invariant interaction
Hamiltonian as well as $S$-matrix elements are identical to their untwisted analogues and hence by doing a scattering experiment it is rather difficult to distinguish between 
a twisted and untwisted theory. Perhaps the best place to look for potential signatures of such particles is to look at the statistical properties and to 
construct observables which depend crucially on the statistics followed by these particles. For example, the thermal correlation functions of the twisted theory turn out to be 
different from that of the untwisted theory and hence the thermodynamic quantities in the two theory will be different. Such differences can probably be detected in appropriate
condense matter systems. Also, we expect that experiments in quantum optics, owing to their crucial dependence on the statistics of the particles involved, can also be able to 
detect such deviations. Furthermore, relaxing the demand of $SU(N)$ invariance leads to differences between the two theories and for certain interaction 
terms the twisted theory turns out to be nonlocal although its analogous untwisted theory is local. Such nonlocal effects can also provide potential signatures
for twisted theory. We plan to address this issue in more details in a future work.
As an interesting application of these ideas we showed that the marginal ($\beta$-)
deformations of the scalar matter sector of $\mathcal{N}=4$ SUSY Hamiltonian density can be described in terms of a twisted interaction Hamiltonian density and hence the twisted internal
symmetries provide a natural framework for the discussion of such theories.  
 
We then constructed more general twisted statistics which can be viewed as internal symmetry analogue of dipole theories. We also discussed the construction of interaction 
terms and scattering formalism for it. The main results for general twists are same as those for the antisymmetric twist. 

We ended the paper with discussion of causality of the twisted field theories. We showed that the twisted fields are noncausal and hence a generic observable constructed out 
of them is also noncausal. Inspite of this it is possible to construct certain interaction Hamiltonians, e.g. the $SU(N)$ invariant interaction Hamiltonian, which are causal and 
satisfy cluster decomposition principle. In view of the nonlocal nature of the twisted fields, these field theories (with appropriate generalizations) have the potential 
to circumvent the Coleman-Mandula no-go theorem \cite{coleman}. We plan to discuss such theories in future works.
Also, in this work we did not discussed the possibility of spontaneous symmetry breaking. Such a scenario is quite interesting but it requires a separate discussion. 
We plan to discuss it also in a future work.


\section*{Acknowledgements}

RS will like to thank M. Paranjape and Groupe de Physique des Particules of Universit\'e de Montr\'eal, where a major part of the work was done, for their wonderful hospitality. 
SV would like to thank ICTP, Trieste for support during the final stages of this work. RS is supported by C.S.I.R under the award no: F. No 10 – 2(5)/2007(1) E.U. II.  \\



\end{document}